\documentclass[twocolumn,showpacs,preprintnumbers,amsmath,amssymb
 aps,
 prb,
 lengthcheck,%
]{revtex4-1}
\usepackage{amssymb}
\usepackage{graphicx}
\usepackage{dcolumn}
\usepackage{bm}
\usepackage{hyperref}
\usepackage{makeidx}

\usepackage[english]{babel} 
\usepackage[utf8]{inputenc}
\usepackage{color}
\usepackage{graphicx}
\usepackage{amsmath}

\usepackage{latexsym}
\usepackage{amsthm}

\usepackage{bbm}
\usepackage{float}
\usepackage{mathtools}

\makeindex
\def\>{\right\rangle}
\def\<{\left\langle}
\def\be{\begin{equation}}
\def\ee{\end{equation}}
\def\ba{\begin{array}{l}}
\def\ea{\end{array}}
\def\beq{\begin{eqnarray}}
\def\eeq{\end{eqnarray}}

\begin{document}

 
\title{Unified scattering approach to Josephson current and thermal noise in BCS and topological superconducting junctions}

\author{R. Jacquet,$^1$ J. Rech,$^1$ T. Jonckheere,$^1$ A. Zazunov,$^2$ and T. Martin$^1$}
\affiliation{$^1$ Aix-Marseille Universit\'e, Universit\'e de Toulon, CNRS, CPT, UMR 7332, 13288 Marseille, France}
\affiliation{$^2$ Institut f\"ur Theoretische Physik IV, Heinrich Heine Universit\"at, D-40225 Düsseldorf, Germany}
\date{\today}

\begin{abstract}
We present a unified description of a junction between $s$-wave (BCS) superconductors and a junction between $p$-wave superconductors in a topologically nontrivial phase, which relies on a scattering state expansion. We compute Josephson current and thermal noise in the two kinds of junction and exhibit some characteristic features for a junction of two topological superconductors hosting Majorana zero-energy modes.
\end{abstract}


\maketitle

\section{Introduction}

The idea for the possible existence of Majorana fermions, particles which are their own antiparticles~\cite{majorana}, originates in particle physics. Despite some candidates, they have not been confirmed as elementary particles yet~\cite{rmp87_elliott}. Nevertheless, it has been realized that they can occur as collective excitations in many-body electronic systems~\cite{prb61_read}. The lattice toy model of a spinless chain with $p$-wave pairing proposed by Kitaev~\cite{kitaev01} exhibits two topologically distinct phases, one of which supporting zero-energy end states with Majorana properties. The proposals for practical realizations of topologically nontrivial phases hosting Majorana fermions are numerous and followed the original proposal by Fu and Kane~\cite{prl100_fu_kane} whose idea was to circumvent the problem of knowing whether the elusive $p$-wave superconductivity does exist or not in Nature by instead using proximity effect with an $s$-wave superconductor. At the view of the growing interest rising in the condensed matter community, some dedicated reviews have been published~\cite{alicea_rev,beenakker_rev,flensberg_rev}.
 
In the context of condensed matter physics, the motivation for the search of Majorana fermions goes beyond the intriguing new fundamental physics that these exotic particles will inevitably provide if they are identified. The exchange of Majorana fermions bound to topological defects in a 2D system results in non-Abelian statistics~\cite{prb61_read,prl86_ivanov} so that Majorana qubits can be used as a fault-tolerant building block for a universal quantum computer~\cite{rmp80_nayak,ann_phys_kitaev}. Indeed, the nonlocal storage of the quantum information in such qubits adresses in an alternative way the issue of decoherence that practical implementations face. 

Among all solid-state settings for the synthesis of Majorana fermions as emergent quasiparticle excitations, semiconductor-superconductor hybrid systems~\cite{prl104_sau,prb81_alicea} and especially proposals based on 1D semiconducting wires\cite{prl105_lutchyn,prl105_oreg}, the so-called Majorana wires, have received a lot of attention. Indeed, spinless $p$-wave superconductivity can be mimicked with materials which are commonly used in nanofabrication. Strong Rashba spin-orbit coupling together with a Zeeman splitting for time-reversal symmetry breaking are the key ingredients to create an isolated spin band inside which superconductivity can be proximity-induced using a conventional $s$-wave superconductor. Moreover, braiding of non-Abelian anyons can be achieved in an effective 2D network of Majorana wires~\cite{nphys_alicea}. 

Considering the relative ease to design Majorana wires and their versatility for future quantum computation schemes, several experiments have implemented these proposals and have quite rapidly reported a zero-bias conductance peak~\cite{science336_mourik,nat_phys_das,prb87_churchill,nano_letters_deng,prl110_finck} compatible with the presence of Majorana zero-energy modes. Fractional $4\pi$ periodic Josephson current~\cite{kitaev01,prl100_fu_kane,prb79_fu_kane} has also been observed~\cite{nature_rokhinson} but does not give decisive evidence for Majorana fermions neither. Potential non-topological origins for such signatures have to be ruled out. In particular, discriminating features in conductance measurements have motivated a lot of works. For example, recent theoretical predictions~\cite{prb97_sau,prb97_flensberg} have followed the experimental study in Ref.~\onlinecite{nature_albrecht} which has reported exponential suppression of energy splitting with wire length compatible with the hybridization of Majorana end states in a Coulomb blockaded nanowire. At the cutting edge of experimental progress in the search for Majorana fermions, a proper quantization of zero-bias conductance has recently been reported~\cite{prl119_nichele,nature556_zhang}.

Non-Abelian exchange statistics would be the ultimate compelling proof for the observation of Majorana fermions. Before the implementation of T-junctions~\cite{nphys_alicea} for braiding experiments, quantum transport in systems supporting Majorana states can be investigated in the aim to propose other kinds of signature for the presence of Majorana fermions. The study of a junction between two $p$-wave superconductors in a topologically nontrivial phase (TS-TS junction) is motivated by the inherent presence of Majorana fermions at the interface~\cite{kitaev01}. 

Short BCS superconducting constrictions (junctions in which the contact length is much shorter than the superconducting coherence length) have extensively been studied in the litterature, including computations of Josephson current~\cite{phyb165,prl66_beenakker,prl67_beenakker} and thermal noise~\cite{prb53_martin-rodero}. Quasiclassical helical mode expansion and Dirac potential modelization for backscattering at the contact point between wires~\cite{prb48_shumeiko} are well suited to describe a junction between two BCS superconductors (S-S junction). This framework has been adopted to adress decoherence issues of the so-called Andreev level qubit~\cite{prl90_zazunov}, including interactions with acoustic phonons~\cite{prb71_zazunov} and quasiparticle poisoning~\cite{prb89_olivares,prb90_zazunov,condmat_riwar}. The complete description in terms of scattering eigenstates has been provided in the form of bispinors~\cite{prb89_olivares,prb90_zazunov} (which collect right or left movers with spin up or down). A very close description for a (chiral) TS-TS junction using (constrained) bispinors is possible. 

In this work, we investigate a unified treatment of both one-dimensional short S-S and TS-TS junctions, and we focus on the computation of the electronic current statistics at equilibrium: the Josephson current and the thermal noise. Our calculations are based on the matching of the scattering states of the Bogoliubov - de Gennes equation. While Green's function based theoretical frameworks (e.g. Ref.~\onlinecite{prb94_zazunov}) have been used to study TS-TS junctions, our approach has the advantage that TS-TS junctions and S-S junctions can be treated in parallel, which offers a global picture of both type of junctions.


The paper is organized as follows. In Sec.~\ref{sec_bdg}, we introduce the Bogoliubov - de Gennes (BdG) Hamiltonian which describes either a BCS or a topological superconducting wire as well as the junction backscattering modelization leading to a matching condition for scattering eigenstates which encodes the nature of the superconductors in contact. Sec.~\ref{sec_majo} is devoted to the emergence of zero-energy bound states in a TS-TS junction and their Majorana properties. Then, we introduce in Sec.~\ref{sec_current} the current operator and the statistics we want to investigate. The Andreev sector is studied in Sec.~\ref{sec_noise_aa}. Transitions involving continuum states give rise to a non-resonant noise that we calculate in Sec.~\ref{sec_noise_c}. Finally, we give some concluding remarks in Sec.~\ref{sec_conclusion}. 

We consider units in which Planck and Boltzmann constants, together with the elementary charge are unity, \textit{i.e.} $\hbar=1$, $k_B=1$, $e=1$.

\section{Unified description of S-S and TS-TS junctions in terms of BdG scattering eigenstates}\label{sec_bdg}

We consider first a one-dimensional BCS superconductor with Fermi energy $E_F=k_F^2/(2m)$, gap energy $\Delta$ and position dependent superconducting phase $\phi(x)$. In the quasiclassical approximation~\cite{prb48_shumeiko}, the fermionic annihilation operator $\psi_\sigma(x)$, for a particle with spin $\sigma=\uparrow,\downarrow$ at position $x$, is expanded around Fermi points (with momenta $\pm k_F$), introducing slowly varying (\textit{i.e} varying on a scale much larger than $\lambda_F=2\pi/k_F$) envelope operators $\psi_{j\sigma}(x)$ ($j=R,L$), according to
\be
\psi_\sigma(x)=\text{e}^{ik_Fx}\psi_{R\sigma}(x)+\text{e}^{-ik_Fx}\psi_{L\sigma}(x)~.
\ee
We define the following Nambu spinors which collect these right and left movers (with Fermi velocity $v_F=\sqrt{2E_F/m}$)   
\be
\psi_+=
\left(
\begin{matrix}
\psi_{R\uparrow}\\
\psi_{L\downarrow}^\dag
\end{matrix}
\right)\quad\text{and}\quad
\psi_-=
\left(
\begin{matrix}
\psi_{L\uparrow}\\
\psi_{R\downarrow}^\dag
\end{matrix}
\right)~.
\ee
These spinors have given $\pm$ helicity and are combined into the bispinor
\be
\Psi=\left(
\begin{matrix}
\psi_+\\
\psi_-
\end{matrix}
\right)~.
\ee
We define the Pauli matrices $\tau_i$ and $\sigma_i$ ($i=x,y,z)$ which act respectively in Nambu and right/left-mover spaces. Then, the Hamiltonian can be written as
\be
H_\text{BCS}=\int\text{d}x\,\Psi^\dag(x)\left[-iv_F\partial_x\,\sigma_z\tau_z+\Delta\tau_x\,\text{e}^{i\phi(x)\tau_z}\right]\Psi(x)~.
\ee
If the phase is homogeneous accross the material, that is $\phi(x)=\varphi$, we can gauge it out with the substitution $\Psi\leftarrow\text{e}^{i\frac{\varphi}{2}\tau_z}\Psi$, so that we obtain 
\be
H_\text{BCS}=\int\text{d}x\,\Psi^\dag(x)\,{\cal H}_\text{BdG}\left(\partial_x\right)\Psi(x)~,\label{eq_ham_bcs}
\ee
where the BdG Hamiltonian reads
\be
{\cal H}_\text{BdG}\left(\partial_x\right)=-iv_F\partial_x\,\sigma_z\tau_z+\Delta\tau_x~.
\ee

Setting the chemical potential to zero in the Kitaev chain model~\cite{kitaev01} leads to a topologically nontrivial phase which supports zero-energy Majorana end states. A linearization around Fermi points yields the following Hamiltonian for a topological superconducting wire (TS wire)~\cite{prb94_zazunov} 
\be
H_\text{TS}=\sum_q\Phi_q^\dag\left[v_Fq\,\tau_z+\Delta\,\text{e}^{i\varphi\tau_z}\tau_x\right]\Phi_q\quad\text{with}\quad
\Phi_q=
\left(
\begin{matrix}
c_{k_F+q}\\
c_{-k_F-q}^\dag
\end{matrix}
\right)~.
\ee
We introduce right and left movers $\psi_j$ ($j=R,L$) by the following Fourier transforms ($l$ is the wire length)
\be
\left\{
\begin{aligned}
&c_{k_F+q}=\int\frac{\text{d}x}{\sqrt{l}}\,\text{e}^{-i(k_F+q)x}\psi_R(x)~,\\
&c_{-k_F-q}=\int\frac{\text{d}x}{\sqrt{l}}\,\text{e}^{i(k_F+q)x}\psi_L(x)~,
\end{aligned}
\right.
\ee
and collect them into the Nambu spinors 
\be
\psi=
\left(
\begin{matrix}
\psi_R\\
\psi_L^\dag
\end{matrix}
\right)~,\quad
\overline{\psi}=i\tau_y\psi^\ast=\left(
\begin{matrix}
\psi_L\\
-\psi_R^\dag
\end{matrix}
\right)~.
\ee
The $^\ast$ operation on a vector consists in the hermitian conjugation of its components. Then, if we define the bispinor
\be
\underline{\Psi}=\frac{1}{\sqrt{2}}\left(
\begin{matrix}
\psi\\
\overline{\psi}
\end{matrix}
\right)~,
\ee
and gauge out the homogenous phase $\underline{\Psi}\leftarrow\text{e}^{i\frac{\varphi}{2}\tau_z}\underline{\Psi}$, we can write the TS wire Hamiltonian as
\begin{equation}
H_\text{TS}=\int\text{d}x\,\underline{\Psi}^\dag(x)\,{\cal H}_\text{BdG}\left(\partial_x\right)\underline{\Psi}(x)~.
\end{equation}
An important distinction with the BCS wire Hamiltonian~\eqref{eq_ham_bcs} resides in the reality constraint $[C\underline{\Psi}]^\ast=\underline{\Psi}$ with $C=\sigma_y\tau_y$ that the bispinor $\underline{\Psi}$ describing a TS wire must obey. In the aim to provide a unified description of BCS and TS wires, we have artificially doubled the number of degrees of freedom for a TS wire and this double counting issue has to be carefully considered when calculating expectation values. 

Expanding the bispinor $\Psi(x)$ (or $\underline{\Psi}(x)$) as $\sum_\nu\chi_\nu(x)c_\nu$ where $c_\nu$ is a fermionic annihilation operator for a quasiparticle in the energy state $E_\nu$, the diagonalization of the Hamiltonian $H_\text{BCS}$ (or $H_\text{TS}$) as $\sum_\nu E_\nu c_\nu^\dag c_\nu$ is effective if the wavefunctions $\chi_\nu$ are eigenstates of the BdG equation:
\begin{subequations}
\begin{align}
\left\{{\cal H}_\text{BdG}(\partial_x)-E_\nu\right\}\chi_\nu(x)=0~,
\\
\int\text{d}x\,\chi_\mu^\dag(x)\chi_\nu(x)=\delta_{\mu\nu}~.
\end{align}
\end{subequations}
There are continuum plane wave states proportional to $\text{e}^{\pm ik_Ex}$ with dispersion relation $E^2=\Delta^2+(v_Fk_E)^2$. Such a state can be written as a sum of incoming and outgoing waves and four possibilities of scattering states can be distinguished at a given energy $E$: the incoming wave can be an electron-like or a hole-like excitation coming from the left or from the right. A label $s=1..4$ is used and $p$ stands for the couple $(E,s)$. There also exist subgap bound states proportional to $\text{e}^{-\kappa_Ex}$, $\kappa_E>0$, with dispersion relation $E^2=\Delta^2-(v_F\kappa_E)^2$. The general form for these wavefunctions are given in Appendix~\ref{app_eigen}.

\renewcommand{\arraystretch}{1.5}
\begin{table}
\centering
\begin{tabular}{c|c|c|}
& topological & BCS \\ \hline
$f(\varphi)$ & $\cos\frac{\varphi}{2}$ & $i\sin\frac{\varphi}{2}$\\ \hline
$g(\theta)$ & $\cosh\theta$ & $\sinh\theta$\\ \hline
$E_A(\varphi)\Delta^{-1}$ & $\sqrt{T}\left|\cos\frac{\varphi}{2}\right|$ & $\sqrt{1-T\sin^2\frac{\varphi}{2}}$\\ \hline
$\delta_A(\varphi)\Delta^{-1}$ & $\sqrt{T}\,\text{sign}\left[\cos\frac{\varphi}{2}\right]\sin\frac{\varphi}{2}$ & $\frac{T}{2}\,\dfrac{\sin\varphi}{\sqrt{1-T\sin^2\frac{\varphi}{2}}}$\\ \hline
$\kappa(\varphi)\xi_0$ & $\sqrt{1-T\cos^2\frac{\varphi}{2}}$ & $\sqrt{T}\left|\sin\frac{\varphi}{2}\right|$\\ \hline
\end{tabular}\caption{Case-dependent functions $f$ and $g$ and expressions of Andreev energy $E_A$ and related quantities.}\label{table_case_func}
\end{table}
\renewcommand{\arraystretch}{1}
\begin{figure}
\includegraphics[scale=0.6]{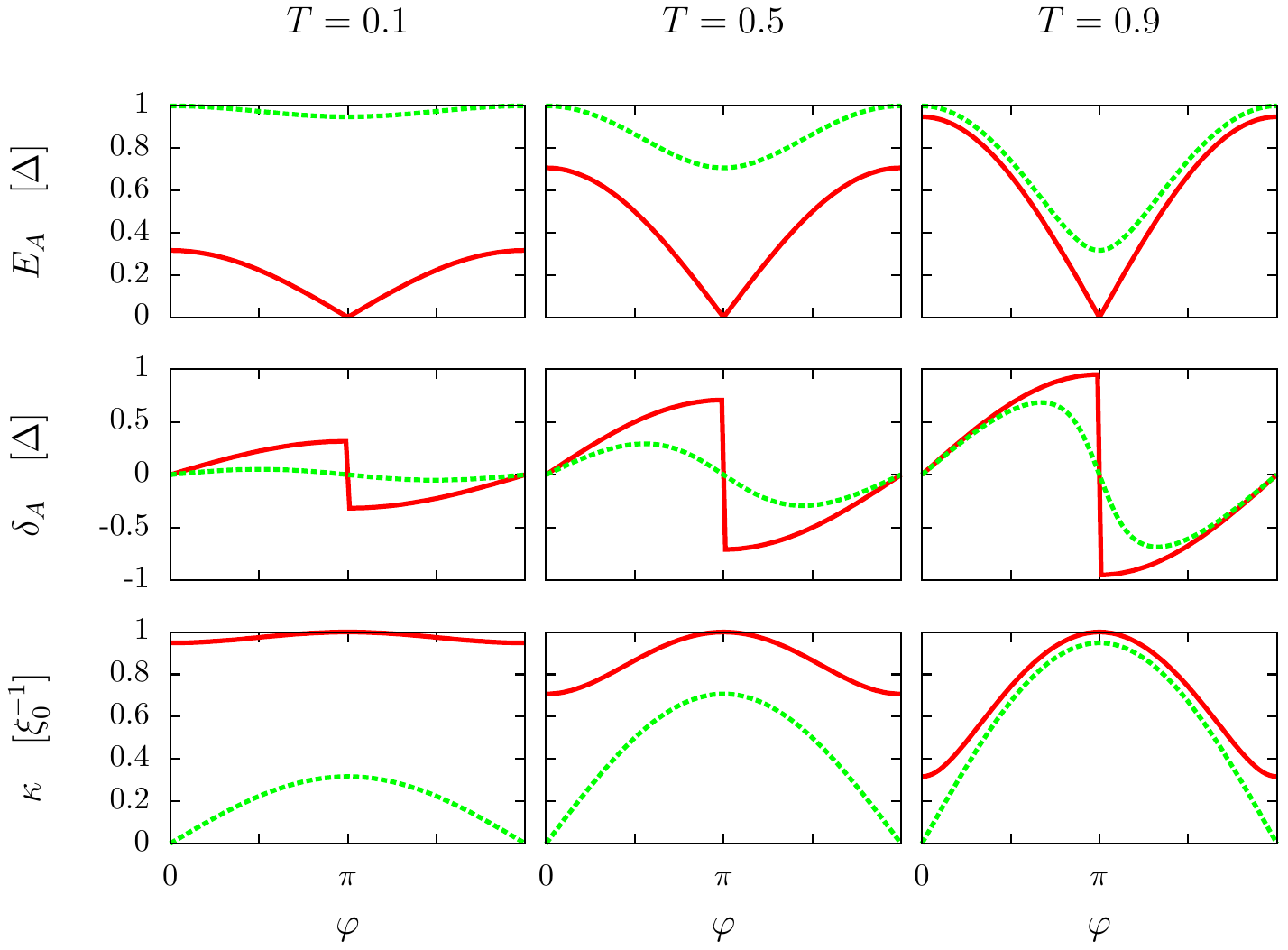}
\caption{Andreev energy $E_A$, its phase derivative $\delta_A$ and damping parameter $\kappa$ in TS-TS (full red) and S-S (dashed green) junctions, and for several transparencies.}\label{fig_andreev_params}
\end{figure}

By introducing a potential barrier that allows backscattering (right mover into a left mover or \textit{vice versa}) in the middle $x=0$ of the 1D superconductor, we modelize a junction between two superconducting wires with identical parameters. While this process is assumed to hold in a spin-preserving way in a S-S junction~\cite{prb48_shumeiko,prb71_zazunov,prb89_olivares,prb90_zazunov,condmat_riwar}, spin flip is necessarily required in a TS-TS junction as a consequence of spin-momentum locking (only one helicity is present). If described by a Dirac delta function, this barrier yields the following matching condition for the BdG eigenstates entering the mode expansion
\be
\chi_\nu(0^-)=\check{T}_\star\chi_\nu(0^+)~,\quad \check{T}_\star=\frac{\text{e}^{i\frac{\varphi}{2}\tau_z}}{\sqrt{T}}\left(1+\sqrt{1-T}\sigma_x\tau_\star\right)~,
\label{eq_matching}
\ee
where $\tau_\star=\tau_z$ in a TS-TS junction and $\tau_\star=1$ in a S-S junction. Because of this single difference between the two cases of junction under study, a description using a few case-dependent functions is possible. Indeed, all results presented in this article only depend on the two functions $f$ and $g$ which are given in Table~\ref{table_case_func}. Remark that, in the limit of perfect transparency $T\to1$, the matching equation is the same in the two cases so that the BdG solutions must coincide. A first consequence of the matching condition is the quantization of subgap states according to $E_\sigma=\sigma E_A$ with $\sigma=\pm$ and
\be
E_A(\varphi)=\Delta\sqrt{\cos^2\frac{\varphi}{2}-(1-T)f^2(\varphi)}~.\label{eq_andreev_energy}
\ee
In Table~\ref{table_case_func}, we give the expressions of the Andreev energy $E_A$, of its phase derivative
\be
\delta_A(\varphi)=-\frac{\partial E_A}{\partial(\varphi/2)}=\frac{T\Delta^2\sin\varphi}{2E_A(\varphi)}~,\label{eq_andreev_delta}
\ee
and of the damping parameter ($\xi_0=\frac{v_F}{\Delta}$ is the superconducting coherence length)
\be
\kappa(\varphi)=\xi_0^{-1}\sqrt{\sin^2\frac{\varphi}{2}+(1-T)f^2(\varphi)}~,\label{eq_andreev_kappa}
\ee
the inverse of which, refered in the following as damping length, gives the extension of the bound states (larger than $\xi_0$). These quantities are displayed in Fig.~\ref{fig_andreev_params}. Andreev energy and damping length are smaller in the topological case than in the BCS one. Consequently, the Andreev levels in a TS-TS junction reside more deeply inside the gap and their associated spatial wavefunctions are more strongly localized around the junction. In both cases, $\varphi=0$ is the maximum of the Andreev energy and, consequently, a zero of the $\delta_A$ quantity and a maximum of the damping length (zero phase difference is a peculiar point in a S-S junction: Andreev levels are ejected at the boundary of the gap and the damping length diverges). In the BCS case, the Andreev energy has a strictly positive minimum in $\varphi=\pi$ (and consequenlty the $\delta_A$ quantity vanishes). In the topological case, the Andreev energy vanishes for $\varphi=\pi$ (which goes with a discontinuity of the $\delta_A$ quantity). To provide a good decoupling of the Andreev pair from continuum states, large transparency $T\sim1$ and bias operating point $\varphi=\pi$ were envisioned in a S-S junction~\cite{prl90_zazunov}. In a TS-TS junction, a good decoupling can always be achieved (whatever the transparency) around the zero $\varphi=\pi$ of the Andreev energy. Remark that, within our convention, see Eq.~\eqref{eq_andreev_energy}, in a TS-TS junction, the evolution with conserved fermion parity leads to a change from one Andreev level branch to the other ($E_\sigma\to E_{-\sigma}$) through the crossing $\varphi=\pi$, at the origin of fractional ($4\pi$ periodic) Josephson effect~\cite{kitaev01}.

The wavefunctions which respect the matching condition~\eqref{eq_matching} are given in Appendix~\ref{app_eigen}.

\section{Majorana states in a TS-TS junction}\label{sec_majo}

\begin{figure}
\includegraphics[scale=0.6]{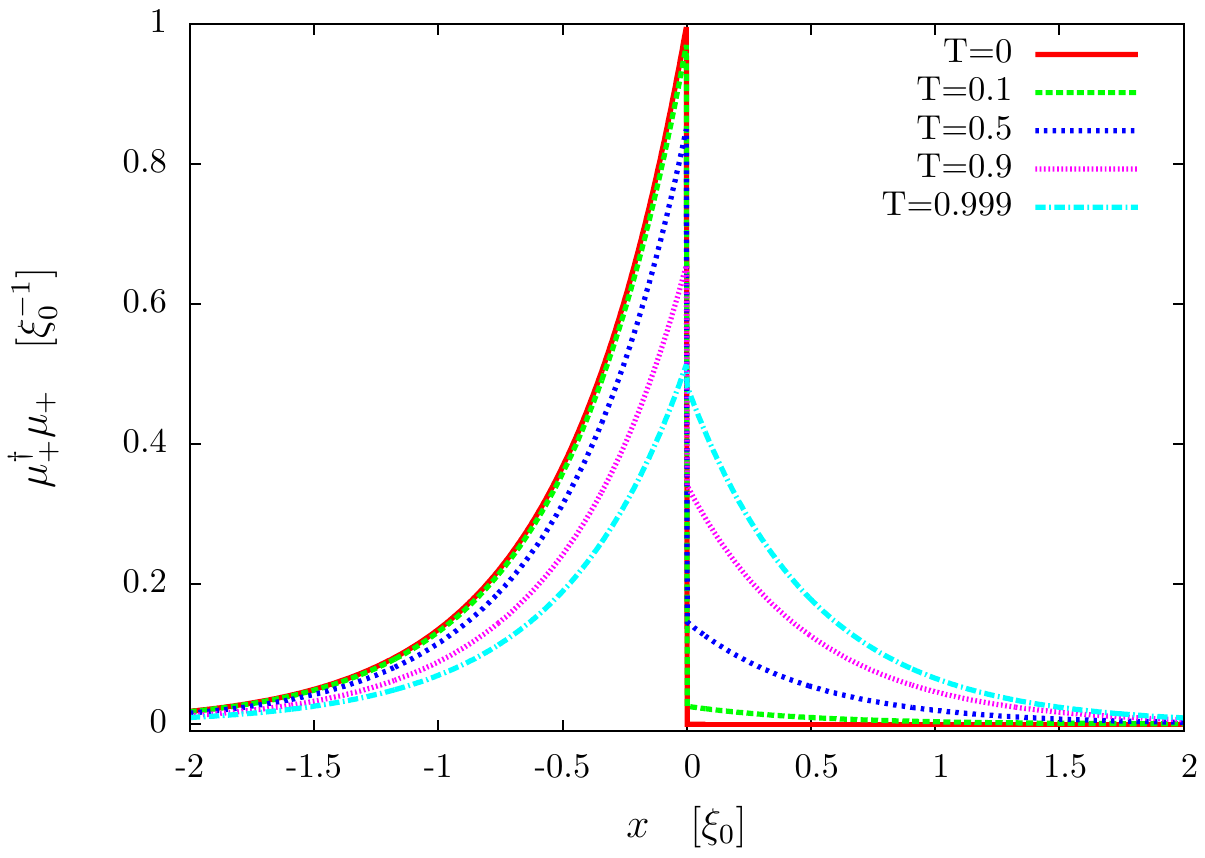}
\caption{Majorana probability density at $\varphi=\pi$ around the junction position $x=0$ for several transparencies.}\label{fig_majo}
\end{figure}

The annihilation operator $c_\sigma$ for the quasiparticle with energy $E_\sigma=\sigma E_A$ is obtained through the projection
\be
c_\sigma=\int\text{d}x\,\chi_\sigma^\dag(x)\underline{\Psi}(x)~.
\ee
Because we have $\left[C\chi_\sigma\right]^\ast=-i\chi_{-\sigma}$ (cf. Appendix~\ref{app_eigen}), the reality constraint $[C\underline{\Psi}]^\ast=\underline{\Psi}$ yields $c_\sigma^\dag=ic_{-\sigma}$. Majorana wavefunctions are obtained as the following superpositions of zero-energy Andreev wavefunctions 
\be
\mu_\sigma(x)=\hat{\chi}_\sigma(x;\varphi=\pi)\quad\text{with}\quad\hat{\chi}_\sigma=\frac{\chi_\sigma-i\,\chi_{-\sigma}}{2}~.
\ee
They verify the reality constraint $\left[C\mu_\sigma\right]^\ast=\mu_\sigma$ so that the associated operators are Majorana fermions
\be
\gamma^\sigma=2\int\text{d}x\,\mu_\sigma^\dag(x)\underline{\Psi}(x;\varphi=\pi)=(\gamma^\sigma)^\dag~,\quad(\gamma^\sigma)^2=1~.
\ee
These operators enter the definition of the Dirac fermion $c_+$ according to
\be
c_+=\frac{\gamma^+-i\gamma^-}{2}~.
\ee
The Majorana wavefunctions read
\be
\mu_\sigma(x)=\frac{\text{e}^{-\frac{|x|}{\xi_0}}}{4\sqrt{\xi_0}}\,m_\sigma(x)\,\text{e}^{i\sigma\,\text{sign}(x)\frac{\pi}{4}\tau_z}
\left(
\begin{matrix}
1\\
\sigma\\
\sigma\\
-1
\end{matrix}
\right)~,
\ee
with 
\be
m_\sigma(x)=\sqrt{1-\sigma\sqrt{T}}-\sigma\,\,\text{sign}(x)\sqrt{1+\sigma\sqrt{T}}~.
\ee
The probability density $\mu_+^\dag\mu_+$ is displayed in Fig.~\ref{fig_majo} ($\mu_-^\dag\mu_-$ is symmetric with respect to the junction position). In the limit $T\to0$, each Majorana wavefunction localizes on a single side of the junction.

\section{Current operator and statistics}\label{sec_current}

Moving to the Heisenberg picture leads to the time dependence for the operators $c_\nu\to c_\nu\text{e}^{-iE_\nu t}$. The current flowing through the junction is given by 
\be
I(t)=\sum_{\nu,\nu'}{\cal I}_{\nu\nu'}c_\nu^\dag c_{\nu'}\text{e}^{i(E_\nu-E_{\nu'})t}
\ee 
where the current matrix elements read
\be
{\cal I}_{\nu\nu'}=v_F\chi_\nu^\dag(0^+)\sigma_z\chi_{\nu'}(0^+)~.
\ee
The equilibrium Josephson current is given by
\be
{\cal I}=\frac{\left\langle I\right\rangle}{d_\star}\quad\text{with}\quad\left\langle I\right\rangle=\sum_\nu{\cal I}_{\nu\nu}n(E_\nu)~,\label{eq_josephson}
\ee
where $n(E)=[1+\exp(\beta E)]^{-1}$ is the Fermi-Dirac distribution at temperature $\beta^{-1}$. $d_\star=1$ in a S-S junction while $d_\star=2$ in a TS-TS junction to avoid double counting. The (unsymmetrized) noise is defined as
\be
S(t,t')=\frac{\left\langle\delta I(t)\delta I(t')\right\rangle}{d_\star}\quad\text{with}\quad\delta I(t)=I(t)-\left\langle I\right\rangle~.
\ee
Its Fourier transform is given by
\be
S(\omega)=\frac{2\pi}{d_\star}\sum_{\nu,\nu'}\left|{\cal I}_{\nu\nu'}\right|^2n(E_\nu)\left[1-n(E_{\nu'})\right]\delta\left[\omega-\left(E_{\nu'}-E_\nu\right)\right]~.\label{eq_noise}
\ee
Zero frequency noise (obtained for $E_\nu=E_{\nu'}$ in the last sum) has a part $S_A(\omega,\varphi)$ due to Andreev bound states that we will briefly discuss. Finite frequency noise can be decomposed as the sum of contributions originated from transitions between Andreev levels (AA), between continuum states (CC) or between an Andreev level and a continuum state (AC), as follows
\be
S(\omega\neq0,\varphi)=S_{AA}(\omega,\varphi)+S_{CC}(\omega,\varphi)+S_{AC}(\omega,\varphi)~.
\ee

\section{Andreev sector}\label{sec_noise_aa}

\begin{figure}
\includegraphics[scale=0.6]{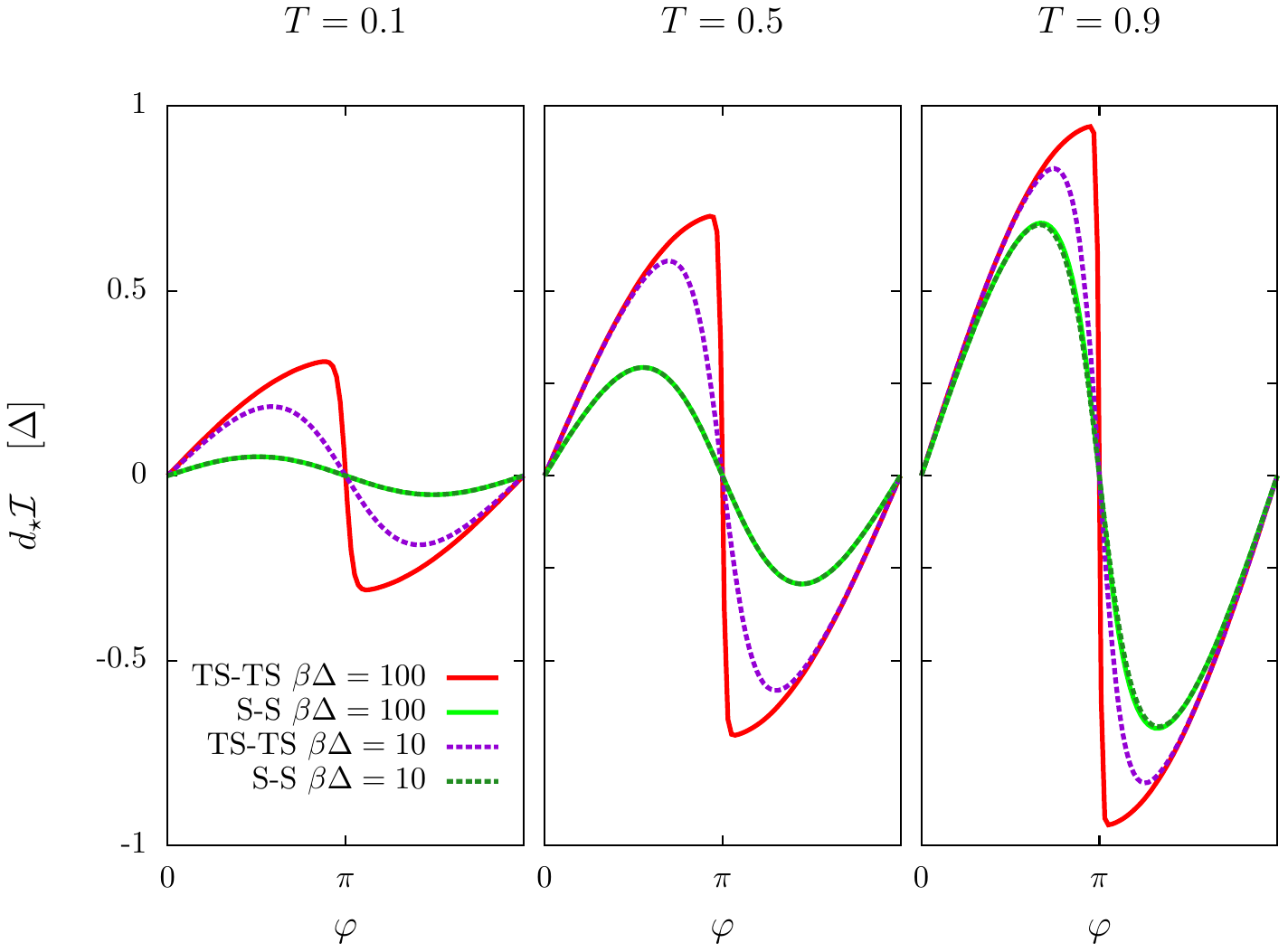}
\caption{Josephson current in TS-TS and S-S junctions for several transparencies and two different temperatures.}\label{fig_josephson}
\end{figure} 
\begin{figure}
\includegraphics[scale=0.6]{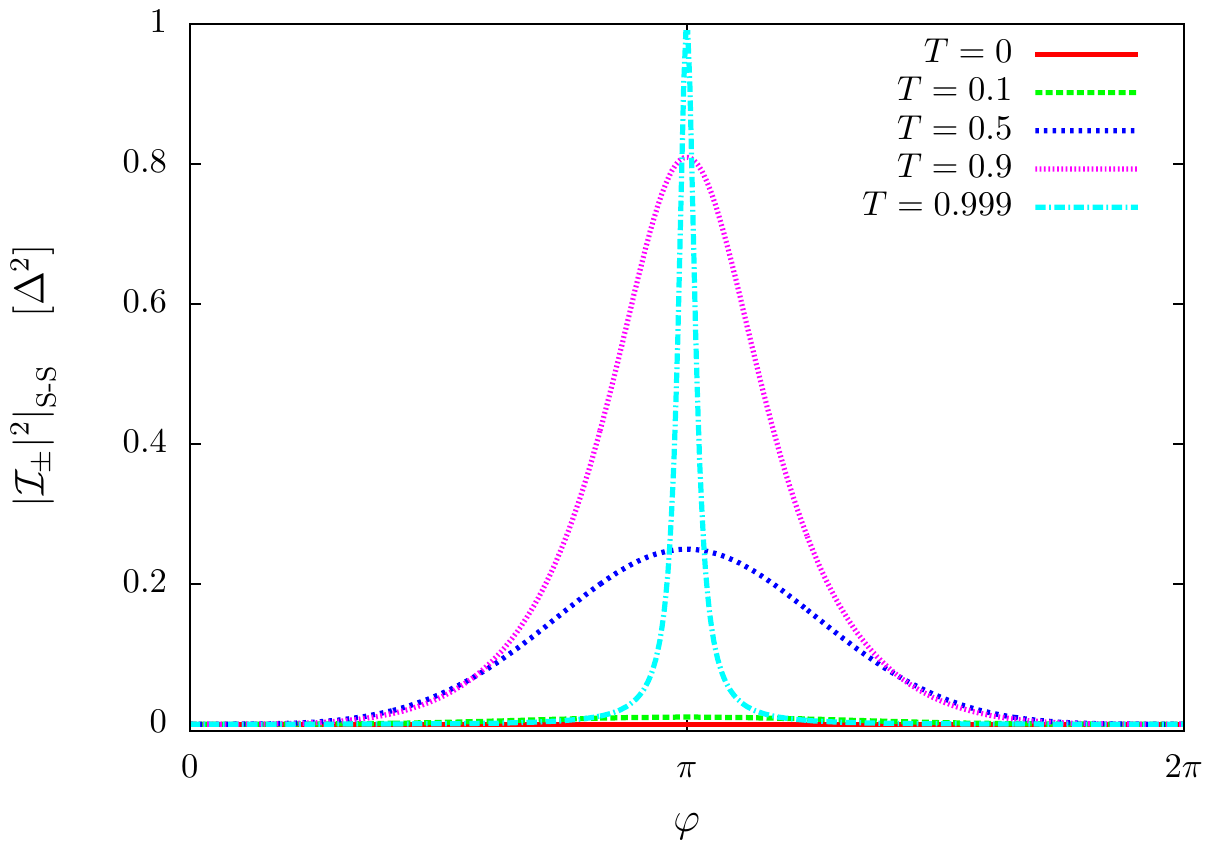}
\caption{Phase dependence of the noise at $\omega=\pm2E_A$ due to transitions between Andreev levels in a S-S junction, for several transparencies.}\label{fig_noise_aa}
\end{figure}

At a given energy $E$, electron-like and hole-like excitations propagating in the same direction carry opposite currents so that the continuum contribution to the Josephson current vanishes
\be
\left(\sum_s{\cal I}_{(E,s)(E,s)}\right)n(E)=0~.
\ee
Then, the Josephson current is entirely carried by Andreev bound states, as expected in short constrictions~\cite{phyb165,prl66_beenakker,prl67_beenakker}, and reads 
\begin{equation}
{\cal I}(\varphi)=\frac{\delta_A(\varphi)}{d_\star}\tanh\frac{\beta E_A(\varphi)}{2}~.
\end{equation}
This result is displayed in Fig.~\ref{fig_josephson} for two different temperatures. For low temperatures, $d_\star{\cal I}$ is given by the function $\delta_A$ (given in Fig.~\ref{fig_andreev_params}) except around the vanishing point of the Andreev energy in a TS-TS junction (for $\varphi=\pi$) where finite temperature effects are noticeable. In the zero-temperature limit, the critical current scales as the ``universal limit'' $\Delta$~\cite{prl66_beenakker,prl67_beenakker}. It is multiplied by $\sqrt{T}/2$ in a TS-TS junction (the $\delta_A$ quantity is an increasing function on the interval $[0,\pi]$ reaching its maximum $\sqrt{T}$ in $\pi$) and by $\sqrt{-T\cos\varphi_0}$ in a S-S junction, where $\varphi_0\in[\pi/2,\pi]$ is the position of the inflexion point of the BCS Andreev energy (maximum of the $\delta_A$ quantity). Although the current in a TS-TS junction is $4\pi$ periodic as long as fermion parity is conserved (fractional Josephson effect), the average (over all parity states) results in a $2\pi$ periodicity~\cite{prb79_fu_kane,prb94_zazunov}. 

Restricting to the Andreev subspace, in the basis $\sigma=(+,-)$, the current matrix is given by
\renewcommand{\arraystretch}{1.5}
\begin{table}[H]
\centering
\begin{tabular}{c|c|c|}
& topological & BCS \\ \hline
$\frac{I_A(\varphi)}{\delta_A(\varphi)}$ & $\left(\begin{matrix}-1&0\\0&1\end{matrix}\right)$ & $\left(\begin{matrix}-1&-i\sqrt{1-T}\,\tan\frac{\varphi}{2}\\i\sqrt{1-T}\,\tan\frac{\varphi}{2}&1\end{matrix}\right)$\\ \hline
\end{tabular}\caption{}
\end{table}
\renewcommand{\arraystretch}{1}
\noindent The Josephson current originates in diagonal elements of this matrix. Because current and Hamiltonian eigenstates are different in the BCS case, the currents carried by Andreev levels $\pm\delta_A(\varphi)$ do not coincide with the current operator eigenvalues $\pm T\Delta\sin\frac{\varphi}{2}$~\cite{prb71_zazunov}. Consequently, transitions between BCS Andreev levels give rise to a noise consisting in Dirac delta peaks at frequency $\omega=\pm2E_A$
\begin{align}
S_{AA}(\omega,\varphi)=\frac{2\pi}{d_\star}\left|{\cal I}_{+-}(\varphi)\right|^2\Big\{&\big(1-n[E_A(\varphi)]\big)^2\delta\left[\omega-2E_A(\varphi)\right]\notag\\
+&\big(n[E_A(\varphi)]\big)^2\delta\left[\omega+2E_A(\varphi)\right]\Big\}~.
\end{align}
The (non-vanishing) squared modulus of the out-of-diagonal element reads
\be
\left.\left|{\cal I}_{+-}(\varphi)\right|^2\right|_\text{S-S}=\frac{T^2(1-T)\sin^4\frac{\varphi}{2}}{1-T\sin^2\frac{\varphi}{2}}\Delta^2~.\label{eq_noise_aa_phi}
\ee
This result has been obtained in Ref.~\onlinecite{prb53_martin-rodero} and we display the phase dependence in Fig.~\ref{fig_noise_aa}. Remark that it vanishes for $\varphi=0$ as a consequence of the expulsion of Andreev levels to the continuum and that it reaches a maximum $T^2\Delta^2$ in $\varphi=\pi$. In the topological case, the current matrix is diagonal in the basis of Andreev states so that there is no contribution to the noise at $\omega=\pm2E_A$: 
\be
\left.S_{AA}(\omega,\varphi)\right|_\text{TS-TS}=0~.
\ee
This is closely related to the emergence of zero-energy Majorana fermions. More precisely, the relation $\left[C\chi_\sigma\right]^\ast\propto\chi_{-\sigma}$ crucial in the construction of Majorana wavefunctions (cf. Sec.~\ref{sec_majo}) and the result ${\cal I}_{+-}=0$ both originate in a single relation between Andreev state coefficients (cf. Appendix~\ref{app_eigen}). Physically, the absence of direct transitions between Andreev levels in the topological case is related to fermion-parity conservation~\cite{prb79_fu_kane}.

Let us conclude our study of the Andreev sector by giving the zero frequency noise
\be
\hspace{-5pt}S_A(\omega,\varphi)=\frac{4\pi\delta_A^2(\varphi)}{d_\star}\,n[E_A(\varphi)]\big(1-n[E_A(\varphi)]\big)\,\delta(\omega)~.\label{eq_zero_freq_noise}
\ee
This result has also been obtained in Ref.~\onlinecite{prb53_martin-rodero} for a S-S junction. In such a junction with transparency ${T<1}$, lowering temperature suppresses this noise resonance since the occupation factor $n(1-n)$ vanishes. For TS-TS junctions, due to fermion parity conservation, Eq.~\eqref{eq_zero_freq_noise} predicts $S_A = 0$ for all $\varphi$ except $\varphi = \pi\,\text{mod} (2\pi)$. At the points of Andreev level crossings the zero-frequency noise~\eqref{eq_zero_freq_noise} may be present at any low temperature provided that the junction is in a mixed parity state, with (temperature independent) $n(0)=1/2$.


\section{Transitions involving continuum states}\label{sec_noise_c}

\begin{figure}
\includegraphics[scale=0.6]{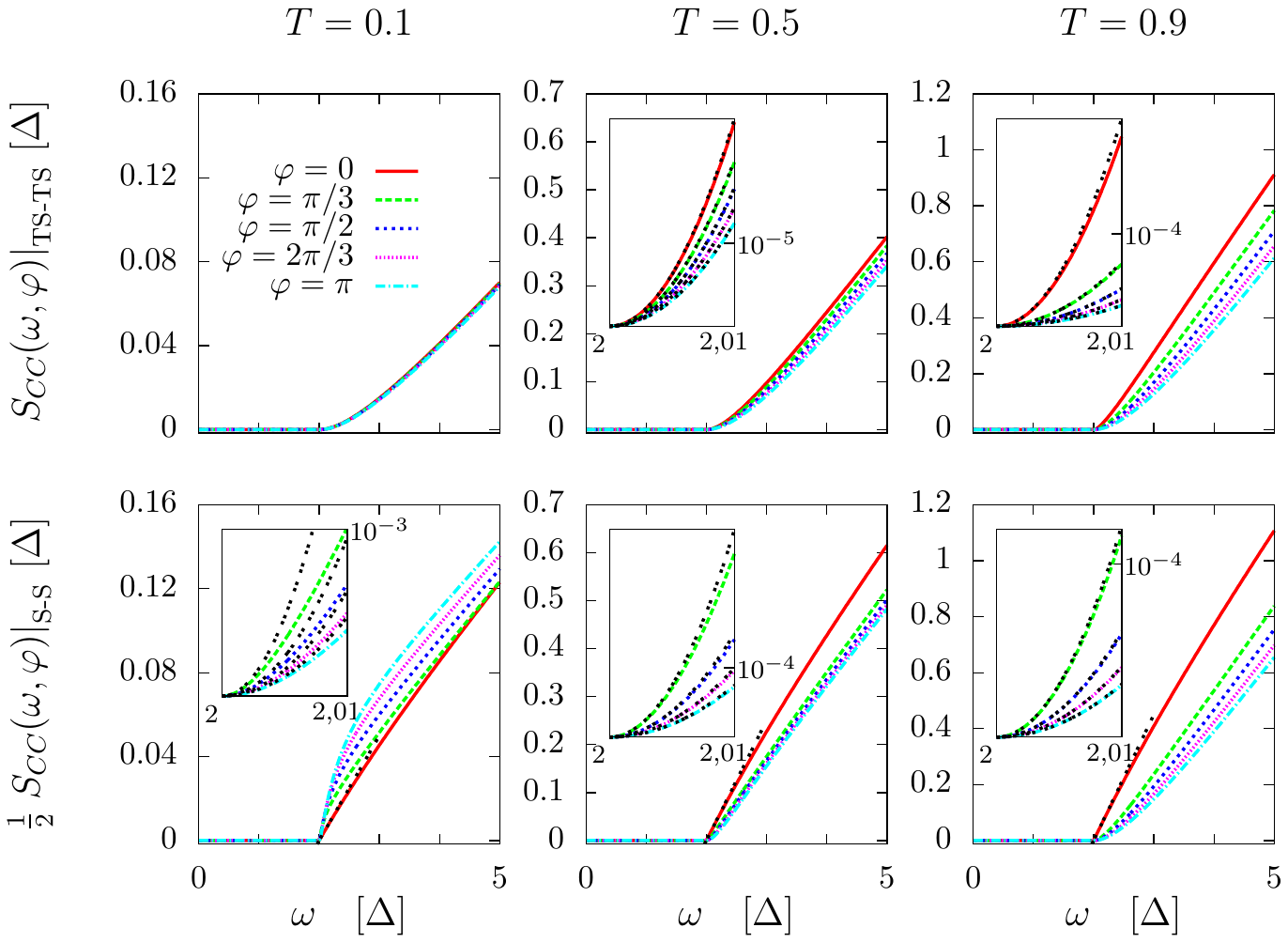}
\caption{Constant $\varphi$ cuts (the key is given in the first panel) of the CC contribution to the noise in TS-TS (top) and S-S (bottom) junctions, and for several transparencies. The insets are enlargements just above the threshold $2\Delta$ and the black lines are the parabolic approximations~\eqref{eq_noise_cc_threshold}. In a S-S junction with $\varphi=0$, the expected linear behaviors~\eqref{eq_noise_cc_threshold_linear} are given by the black lines in the main panels.}\label{fig_noise_cc_cut_phi}
\end{figure}
\begin{figure}
\includegraphics[scale=0.6]{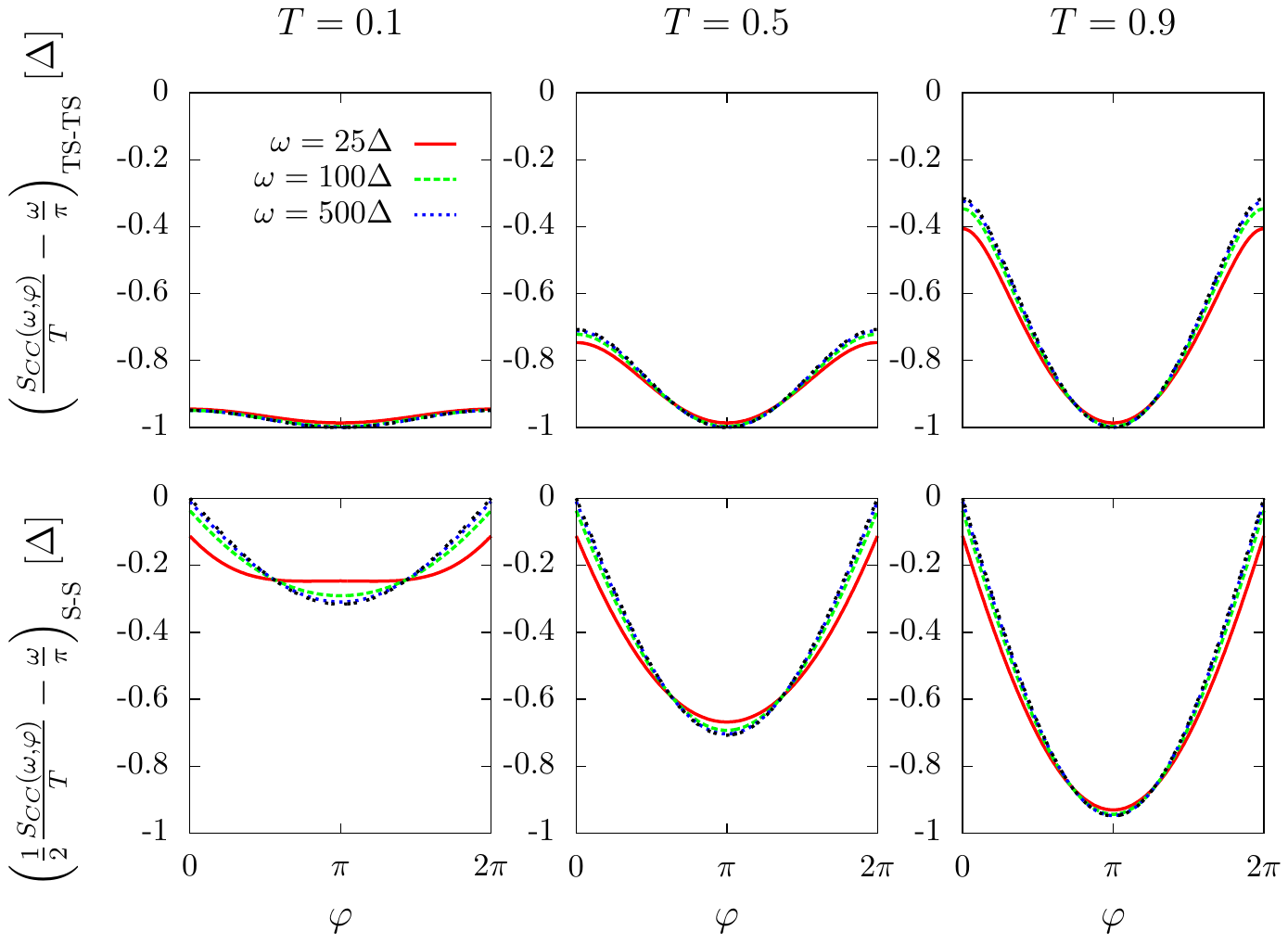}
\caption{Constant $\omega$ cuts (the key is given in the first panel) of the CC contribution to the noise in TS-TS (top) and S\protect\nobreakdash-S (bottom) junctions, and for several transparencies. The noise has been renormalized by the transparency and the linear predominant term at large energies has been substracted: the black lines give the high energy expectation according to Eq.~\eqref{noise_cc_high_energy} which is $-v_F\kappa(\varphi)$.}\label{fig_noise_cc_cut_omega}
\end{figure}

In this section, we provide some analytical results including some interesting limits. The sums over continuum indexes can be decomposed as $\sum_p=\sum_E\sum_s$ and we use the dispersion relation to perform the substitution $\sum_E\to\frac{l}{2\pi v_F}\int\text{d}E\,\Theta\left(|E|-\Delta\right)\frac{|E|}{\sqrt{E^2-\Delta^2}}$. We also give some numerical results obtained for a low temperature $\beta^{-1}=0.01\Delta$ except when specified (during the investigation of finite temperature effects at the end of this section). Remark that, in all figures, noise quantities are multiplied by $d_\star/2$ so that the physical quantities are readily obtained for a TS-TS junction (while for a S-S junction, data are to be doubled).

Calculating the current matrix elements between continuum states, we can write the contribution to the noise due to both intraband and interband transitions as
\begin{spreadlines}{10pt}
\begin{align}
\hspace{-5pt}S_{CC}(\omega,\varphi)=\frac{2}{d_\star}\,\frac{T}{\pi}\int&\text{d}E\,n(E)\left[1-n(\omega+E)\right]\notag\\
&\hspace{-30pt}\times\left[E(\omega+E)+\rho(\varphi)E_A^2(\varphi)\right]\notag\\
&\hspace{-30pt}\times R\left[E,E_A(\varphi)\right]R\left[\omega+E,E_A(\varphi)\right]~.
\label{eq_noise_cc_int}
\end{align}
\end{spreadlines}
The $\rho$ function is defined as
\begin{equation}
\rho=1-(1-\varepsilon)\frac{1-T}{T}\,\frac{\Delta^2-E_A^2}{E_A^2}\label{eq_rhofunc}
\end{equation}
where $\varepsilon=1$ in a TS-TS junction and $\varepsilon=-1$ in a S-S junction (it appears in calculations as $\varepsilon=f(-\varphi)/f(\varphi)=g(-\theta)/g(\theta)$). Notice that $\rho$ is simply 1 in a TS-TS junction. In the integrand~\eqref{eq_noise_cc_int}, the following function is involved
\begin{equation}
R(E_1,E_2)=\Theta\left(|E_1|-\Delta\right)\text{sign}(E_1)\,\frac{\sqrt{E_1^2-\Delta^2}}{E_1^2-E_2^2}~.\label{eq_rfunc}
\end{equation}
It immediately follows that in the zero-temperature limit, as expected, this noise contribution has a $2\Delta$ frequency threshold and 
\be
S_{CC}(\omega,\varphi)\underset{\beta\Delta>\!\!>1}{\sim}\Theta(\omega-2\Delta)S_{CC}^\star(\omega,\varphi)
\ee
where 
\begin{align}
S_{CC}^\star(\omega,\varphi)=\frac{2}{d_\star}\,\frac{T}{\pi}&\int_\Delta^{\omega-\Delta}\text{d}E\,\left[E(\omega-E)-\rho(\varphi)E_A^2(\varphi)\right]\notag\\
&\times\frac{\sqrt{\left[E^2-\Delta^2\right]\left[(\omega-E)^2-\Delta^2\right]}}{\left[E^2-E_A^2(\varphi)\right]\left[(\omega-E)^2-E_A^2(\varphi)\right]}~.
\end{align}
In Fig.~\ref{fig_noise_cc_cut_phi}, constant $\varphi$ cuts are displayed with different energy evolutions depending on the phase difference. Nevertheless, in a TS-TS junction, lowering transparency reduces Andreev energy variations ($E_A(\varphi)\to0$) and the noise is therefore essentially phase independent (constant $\varphi$ cuts collapse on a single curve). The $2\Delta$ threshold is clearly noticeable in Fig.~\ref{fig_noise_cc_cut_phi}. Near the threshold, we have (generally) a parabolic behavior
\be
\frac{d_\star}{2}\,S_{CC}^\star(\omega,\varphi)\underset{\omega/\Delta\to2^+}{\sim}\frac{T\Delta}{4}\,\frac{\Delta^2-\rho(\varphi)E_A^2(\varphi)}{\left[v_F\kappa(\varphi)\right]^4}\,\left(\omega-2\Delta\right)^2~.
\label{eq_noise_cc_threshold}
\ee
In the topological case, it can be further simplified into
\be
\hspace{-5pt}\left.S_{CC}^\star(\omega,\varphi)\right|_\text{TS-TS}\underset{\omega/\Delta\to2^+}{\sim}\frac{T\Delta}{4\,\left[v_F\kappa(\varphi)\right]^2}\,\left(\omega-2\Delta\right)^2~.
\ee
This approximation is compared to the numerical results in the insets of Fig.~\ref{fig_noise_cc_cut_phi}. It is very accurate in a TS-TS junction (except for parameters such that $\kappa\to0$, \textit{i.e.} for $T\to1$ and $\varphi\to0$). In the BCS case, this is not valid for $\varphi=0$ whatever the transparency since the $\kappa$ function vanishes. Instead, we have a linear behavior  
\be
\left.S_{CC}^\star(\omega,\varphi=0)\right|_\text{S-S}\underset{\omega/\Delta\to2^+}{\sim}T\left(\omega-2\Delta\right)~,\label{eq_noise_cc_threshold_linear}
\ee
as also checked in Fig.~\ref{fig_noise_cc_cut_phi}. For a given phase $\varphi\neq0$ in a S-S junction, the parabolic approximation~\eqref{eq_noise_cc_threshold} has a more restricted domain of validity for small transparency since the $\kappa$ function is proportional to $\sqrt{T}$. Whatever the set of parameters such that the approximation~\eqref{eq_noise_cc_threshold} is valid, this is responsible for a smooth cut-off (since there is zero slope at threshold). For large energies, we obtain
\be
\hspace{-5pt}\frac{d_\star}{2}\,S_{CC}^\star(\omega,\varphi)\underset{\omega/\Delta>\!\!>1}{\sim}T\left[\frac{\omega}{\pi}-v_F\kappa(\varphi)\right]+{\cal O}\left(\frac{\Delta^2}{\omega}\right)~.\label{noise_cc_high_energy}
\ee
In Fig.~\ref{fig_noise_cc_cut_omega}, we verify the validity of this expansion by substracting the expected linear term and checking that the result is compatible for sufficiently high energies with the $\kappa(\varphi)$ profile (given in Fig.~\ref{fig_andreev_params}). We see that the convergence is slower in a S-S junction with small transparency as $\kappa$ takes lower values.

\begin{figure}
\includegraphics[scale=0.6]{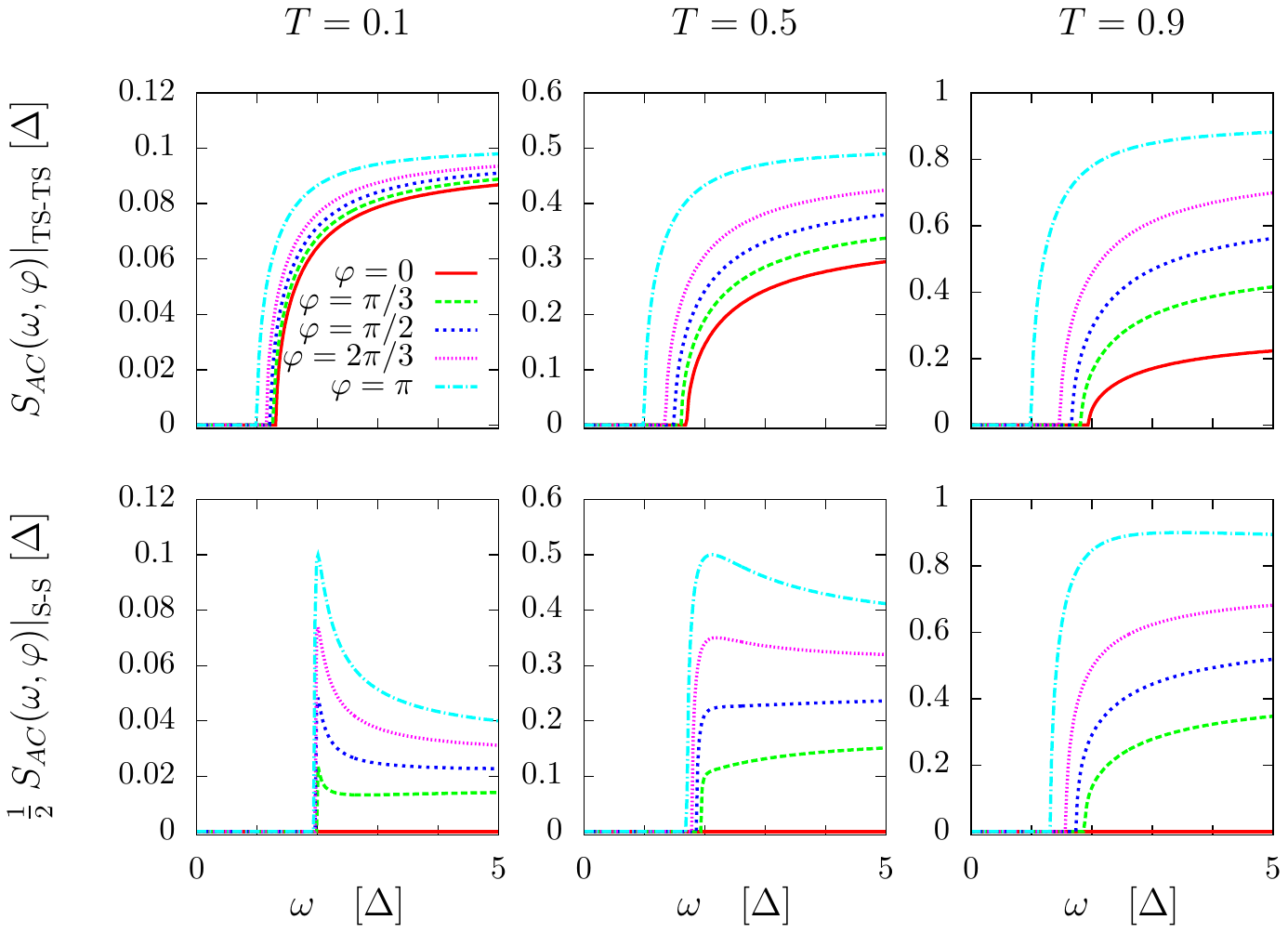}
\caption{Constant $\varphi$ cuts (the key is given in the first panel) of the AC contribution to the noise in TS-TS (top) and S-S (bottom) junctions, and for several transparencies.}\label{fig_noise_ac_cut_phi}
\end{figure}
\begin{figure}
\includegraphics[scale=0.6]{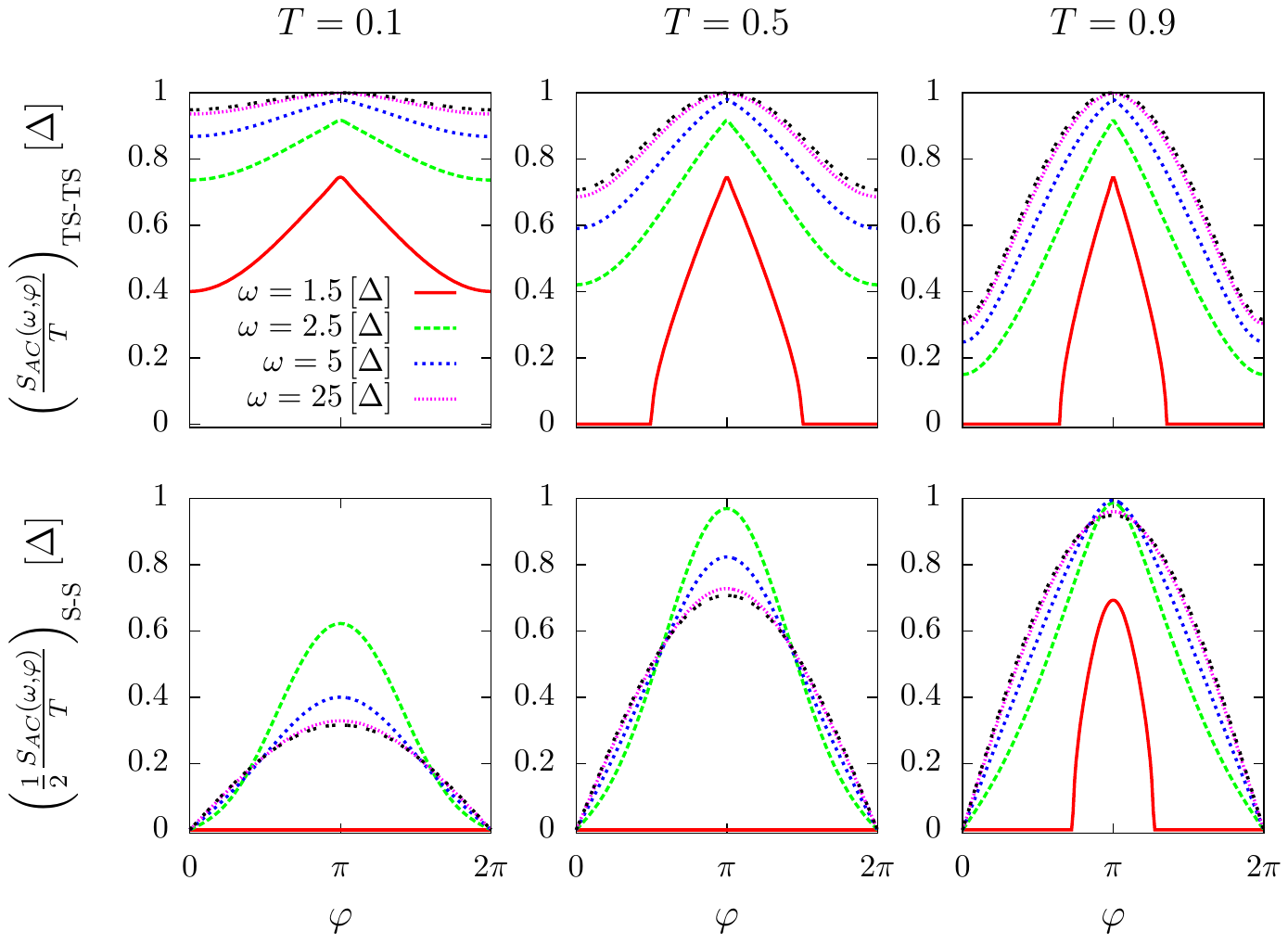}
\caption{Constant $\omega$ cuts (the key is given in the first panel) of the AC contribution to the noise in TS-TS (top) and S-S (bottom) junctions, and for several transparencies. The noise has been renormalized by the transparency. The black lines give the high energy expectation according to Eq.~\eqref{noise_ac_high_energy} which is $v_F\kappa(\varphi)$.}\label{fig_noise_ac_cut_omega}
\end{figure}

Concerning the contribution to the noise due to transitions between an Andreev level and a continuum state, we obtain
\begin{align}
S_{AC}(\omega,\varphi)=\frac{1}{d_\star}\sum_{\sigma=\pm}n&\left[\sigma E_A(\varphi)\right]\big(1-n\left[\omega+\sigma E_A(\varphi)\right]\big)\notag\\
&\times Q_\sigma\left[\omega+\sigma E_A(\varphi),\varphi\right]~.
\end{align}
The function $Q_\sigma$ is given by
\be
Q_\sigma(\omega,\varphi)=2T\,v_F\kappa(\varphi)\left[\omega+\sigma\rho(\varphi)E_A(\varphi)\right]R\left[\omega,E_A(\varphi)\right]~,
\ee
where $\rho$ and $R$ functions have already been introduced in Eqs.~\eqref{eq_rhofunc} and~\eqref{eq_rfunc} respectively. In a S-S junction with zero phase difference, this noise contribution vanishes as the Andreev levels are expelled to the continuum. In the low temperature limit, there is a $\Delta+E_A(\varphi)$ ($\in[\Delta,2\Delta]$) frequency threshold and 
\be
S_{AC}(\omega,\varphi)\underset{\beta E_A(\varphi)>\!\!>1}{\sim}\Theta\left(\omega-[\Delta+E_A(\varphi)]\right)S_{AC}^\star(\omega,\varphi)
\label{eq_noise_ac_low_temp}
\ee
where
\begin{align}
S_{AC}^\star(\omega,\varphi)=\frac{2}{d_\star}\,\frac{T\,v_F\kappa(\varphi)}{\omega}\,&\sqrt{\left[\omega-E_A(\varphi)\right]^2-\Delta^2}\notag\\
&\times\frac{\omega-\left[1+\rho(\varphi)\right]E_A(\varphi)}{\omega-2E_A(\varphi)}~.
\label{eq_noise_ac_low_temp2}
\end{align}
Notice that in the topological case, the last fraction is simply 1. The constant $\varphi$ cuts of Fig.~\ref{fig_noise_ac_cut_phi} show sharp cut-offs (the infinite slope at threshold is due to the square root function in Eq.~\eqref{eq_noise_ac_low_temp2}). On the constant $\omega$ cuts of Fig.~\ref{fig_noise_ac_cut_omega}, the influence of the threshold can be noticed for the value $\omega_0=1.5\Delta\in[\Delta,2\Delta]$: the threshold could be below (first panel) or above (fourth and fifth panels) $\omega_0$ whatever $\varphi$ or the threshold can be crossed for a given $\varphi_0$ according to $\omega_0=\Delta+E_A(\varphi_0)$ (other panels). These constant $\omega$ cuts provide another signature for a TS-TS junction when looking at finite energies $\omega\gtrsim\Delta$ and turning the phase difference across $\varphi=\pi$: there is a cusp reminiscent of the Andreev energy singular vanishing (when considering $E_A(\pi\pm\delta\varphi)=\varepsilon_\pm$ such that $\varepsilon_\pm/\Delta<\!\!<1$ and $\beta\varepsilon_\pm>\!\!>1$, the local slopes on both sides involve the Andreev energy in a linear way). For a S-S junction with transparency ${T<1}$, the Andreev energy has a non-vanishing minimum and we can do the approximation~\eqref{eq_noise_ac_low_temp}-\eqref{eq_noise_ac_low_temp2}. In a TS-TS junction, the Andreev energy vanishes for $\varphi=\pi$ and we get
\begin{equation}
\left.S_{AC}(\omega,\varphi=\pi)\right|_\text{TS-TS}\underset{\beta\Delta>\!\!>1}{\sim}T\Delta\,\frac{\Theta(\omega-\Delta)}{\omega}\sqrt{\omega^2-\Delta^2}~.
\end{equation}
Around the value $\varphi=\pi$, more precisely as long as the Andreev energy $E_A(\varphi)$ is not large compared to the temperature $\beta^{-1}$, we have to conserve the full expression for the low temperature limit
\begin{align}
\hspace{-5pt}
S_{AC}(\omega,\varphi)\underset{\beta\Delta>\!\!>1}{\sim}\sum_{\sigma=\pm}&\Theta\left(\omega-[\Delta-\sigma E_A(\varphi)]\right)\notag\\
&\times n\left[\sigma E_A(\varphi)\right]S_{AC}^{\star\sigma}(\omega,\varphi)~,
\end{align}
where $S_{AC}^{\star-}=S_{AC}^\star$ and $S_{AC}^{\star+}$ is obtained from $S_{AC}^\star$ by substituting $E_A\to-E_A$. Remark that the threshold is then lowered to $\Delta-E_A(\varphi)$. For sufficiently large energies, the $\varphi$ cuts of Fig.~\ref{fig_noise_ac_cut_phi} reach a finite limit which depends on $\varphi$ according to
\begin{equation}
\frac{d_\star}{2}\,S_{AC}(\omega,\varphi)\underset{\substack{\beta\Delta>\!\!>1\\\omega/\Delta>\!\!>1}}{\sim}T\,v_F\kappa(\varphi)+{\cal O}\left(\frac{\Delta^2}{\omega}\right)~.\label{noise_ac_high_energy}
\end{equation}
The constant $\omega$ cuts given in Fig.~\ref{fig_noise_ac_cut_omega} confirm this expectation as they converge to the $\kappa(\varphi)$ profile (given in Fig.~\ref{fig_andreev_params}) if properly renormalized by the transparency.

\begin{figure}
\includegraphics[scale=0.6]{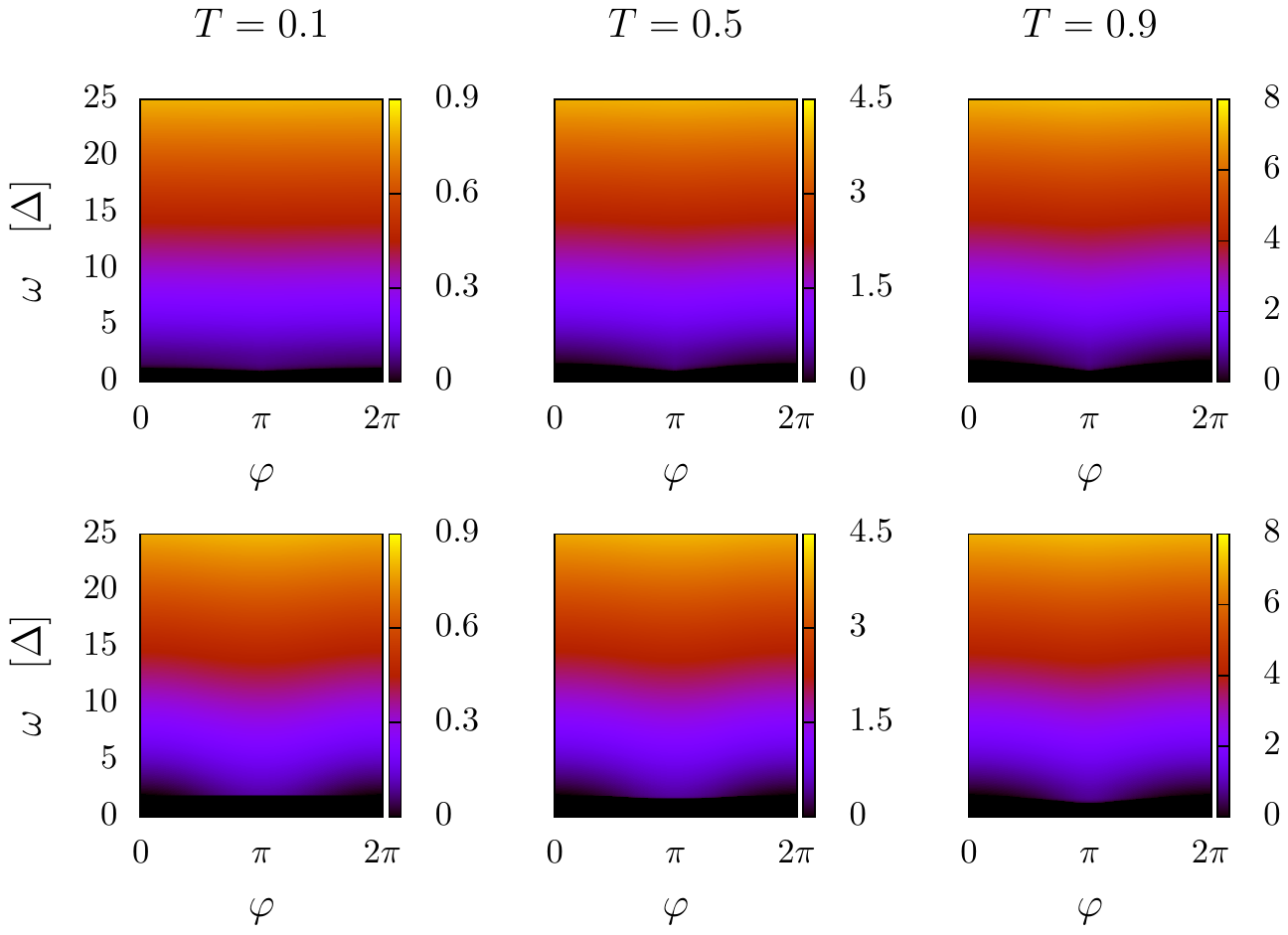}
\caption{Non-resonant noise $\frac{d_\star}{2}S_\text{nr}(\omega,\varphi)$  in TS-TS (top) and S-S (bottom) junctions, and for several transparencies.}\label{fig_noise}
\end{figure}
\begin{figure}
\includegraphics[scale=0.6]{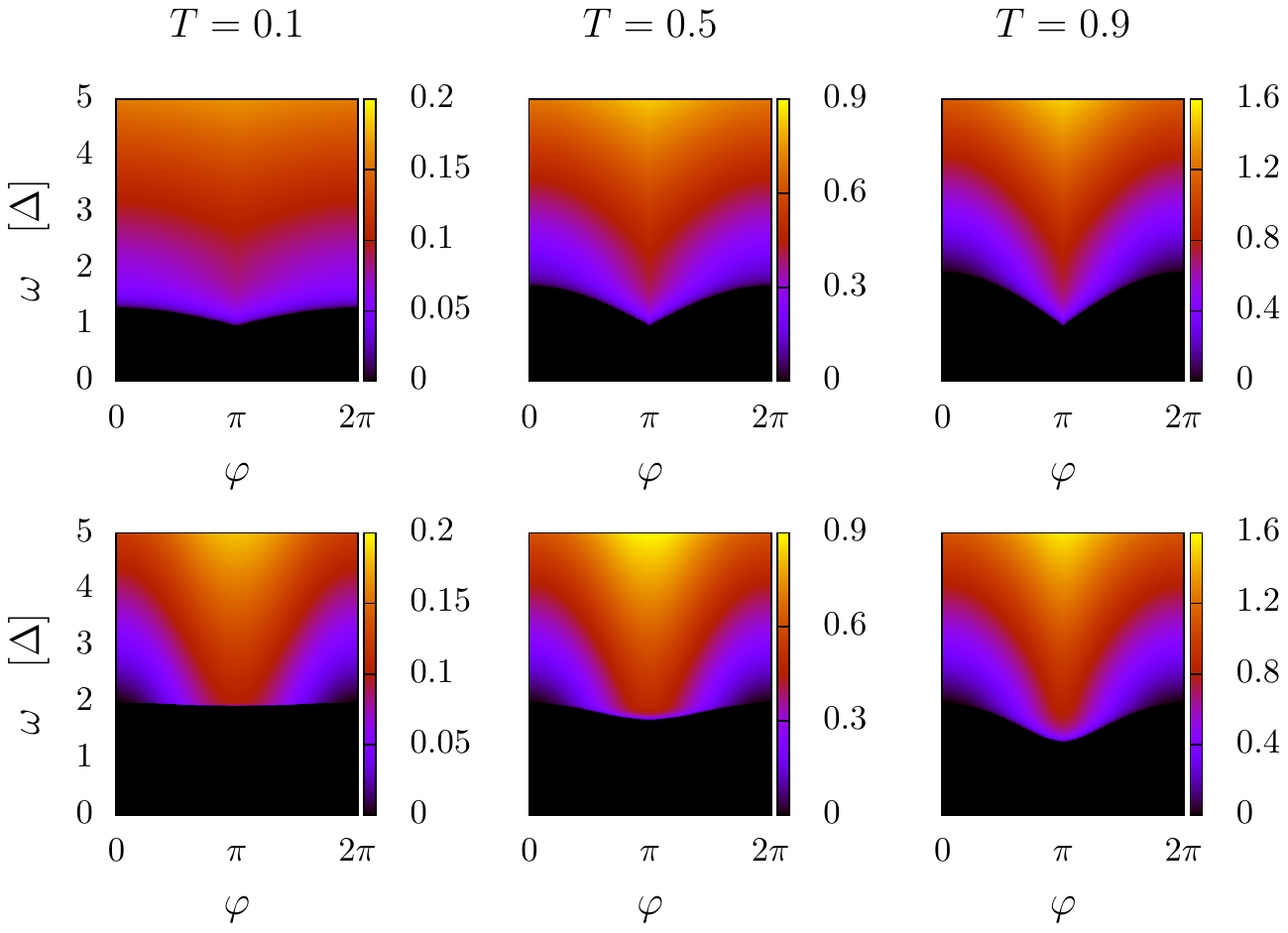}
\caption{Enlargement of Fig.~\ref{fig_noise} around the gap region.}\label{fig_noise_zoom}
\end{figure}
\begin{figure}
\includegraphics[scale=0.6]{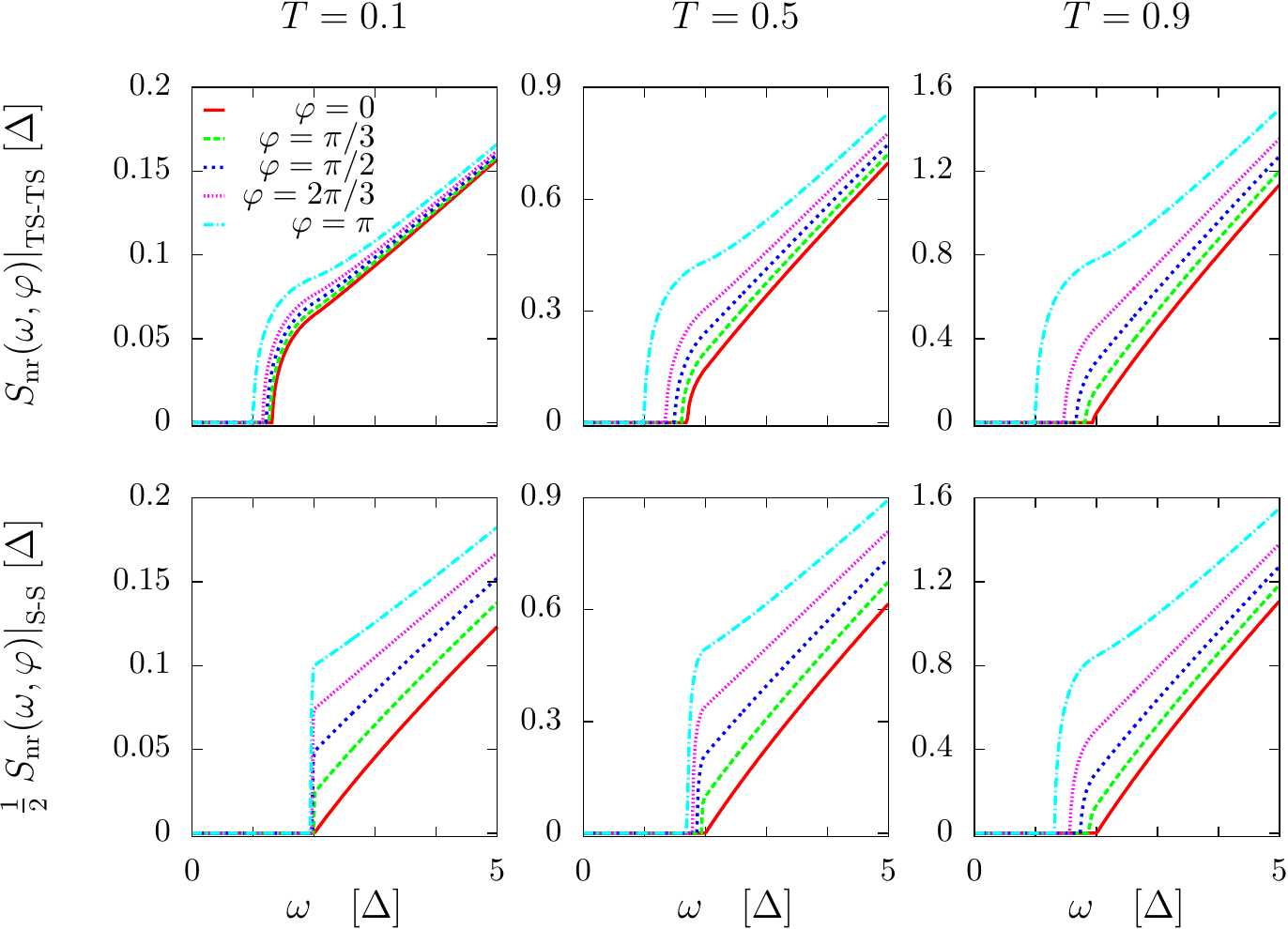}
\caption{Constant $\varphi$ cuts (the key is given in the first panel) of the non-resonant noise in TS-TS (top) and S-S (bottom) junctions, and for several transparencies.}\label{fig_noise_cut_phi}
\end{figure}

Now, we consider the sum of CC and AC contributions which constitutes the non-resonant part of the noise $S_\text{nr}=S_{CC}+S_{AC}$. In a TS-TS junction, this is the total finite frequency noise since transitions inside the Andreev sector are noiseless due to topological protection. In a S\nobreakdash-S junction, two Dirac delta peaks at $\omega=\pm\omega_{AA}$, with $\omega_{AA}=2E_A$, emerge on top of that as a consequence of transitions between Andreev states. 

Let us focus first on the low temperature limit. For energies lower than $\omega_{CC}=2\Delta$ only AC transitions contribute to the non-resonant noise for energies above the threshold $\omega_{AC}=\Delta+E_A$ (this threshold is lowered to $\Delta-E_A$ in the case where the condition ${\beta E_A<\!\!<1}$ is not fulfilled, which occurs around the vanishing point $\varphi=\pi$ of the Andreev energy in a TS-TS junction). For sufficiently high energies, the contribution due to CC transitions predominate and as the result of summing the equivalents~\eqref{noise_cc_high_energy} and~\eqref{noise_ac_high_energy}, we find
\begin{equation}
\frac{d_\star}{2}\,S_\text{nr}(\omega,\varphi)\underset{\substack{\beta\Delta>\!\!>1\\\omega/\Delta>\!\!>1}}{\sim}\frac{T\omega}{\pi}+{\cal O}\left(\frac{\Delta^2}{\omega}\right)~.\label{eq_noisenr_hel}
\end{equation}
In Figs.~\ref{fig_noise} and~\ref{fig_noise_zoom}, the non-resonant noise is given as a map in the $(\varphi,\omega)$ plane. On a large scale of energies, as proposed in Fig.~\ref{fig_noise}, there seems to be little dependence on the phase $\varphi$, while the transparency seems to only affect the magnitude, in agreement with Eq.~\eqref{eq_noisenr_hel}. An enlargement around the gap region is proposed in Fig.~\ref{fig_noise_zoom} where we see a stronger dependence on the $\varphi$ parameter. The (sharp) cut-off reproduces the Andreev energy profile (given in Fig.~\ref{fig_andreev_params}). The transcient regime between AC dominated regime (for $\Delta+E_A(\varphi)<\omega\lesssim2\Delta$) and CC dominated regime (for $\omega>\!\!>2\Delta$) is clearly noticeable on constant $\varphi$ cuts of this map given in Fig.~\ref{fig_noise_cut_phi}, except in a S-S junction with $\varphi=0$ or $T\to0$ and in a TS-TS junction with $\varphi=0$ and $T\to1$ for which the Andreev states are expelled to the continuum (closing the transcient interval). Remark that in a S-S junction with $\varphi=0$, the three frequencies collapse $\omega_{AA}=\omega_{AC}=\omega_{CC}=2\Delta$, both AA and AC noise contributions vanish and rather than inheriting the smooth parabolic cut-off generally obtained for CC transitions, a linear behavior is recovered above the threshold $2\Delta$. Let us emphasize on the increasing behavior of the non-resonant noise as a function of the energy. It is not evident in the BCS case because of the non-monotonic behavior of the contribution due to AC transitions (cf. Fig.~\ref{fig_noise_ac_cut_phi}). Nevertheless, we have no evidence of parameters leading to a decreasing behavior.    

\begin{figure}
\includegraphics[scale=0.6]{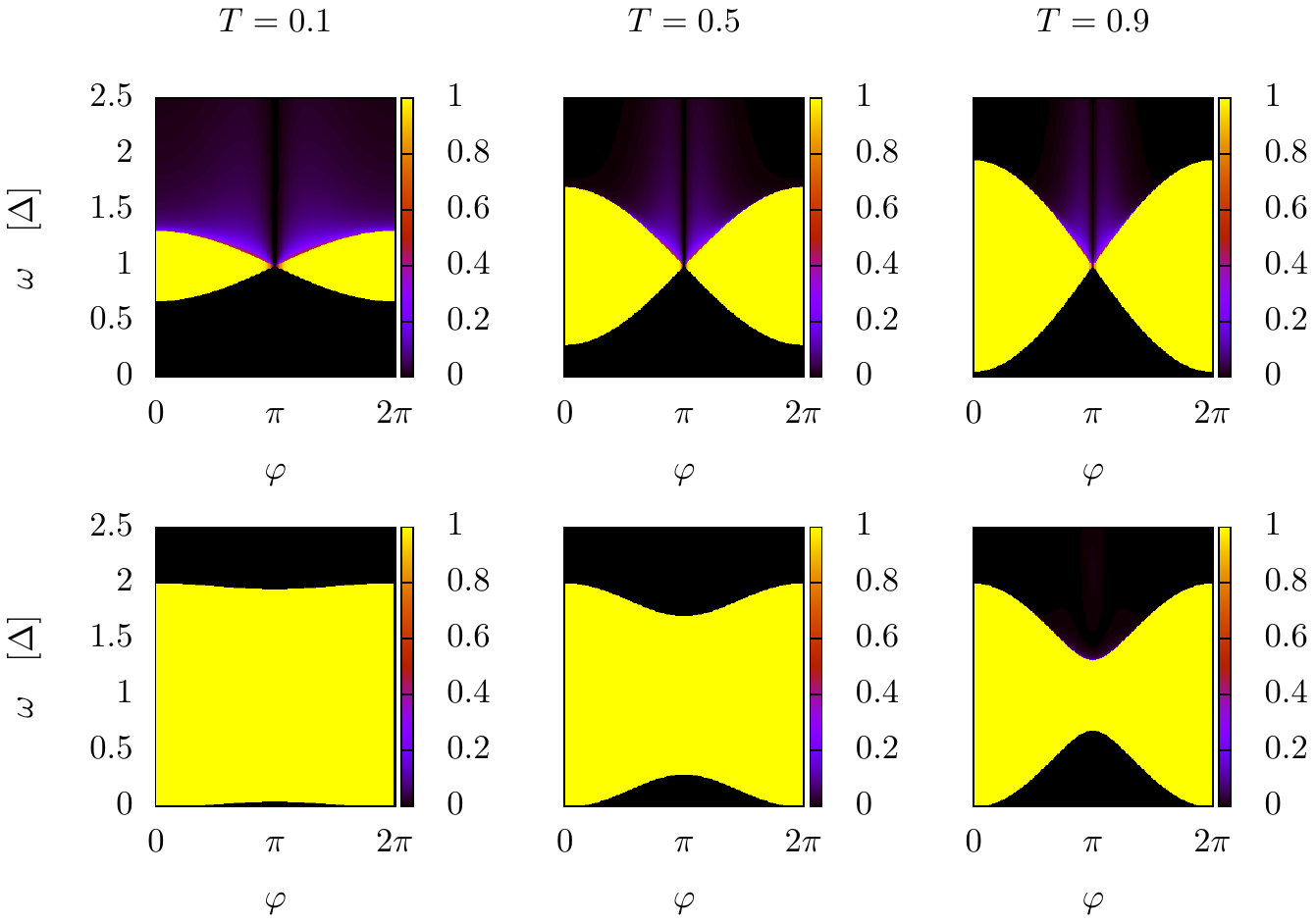}
\caption{Relative difference $|S_\text{nr}^{(1)}-S_\text{nr}^{(2)}|/S_\text{nr}^{(2)}$ between the two non-resonant noises $S_\text{nr}^{(1)}$ and $S_\text{nr}^{(2)}$ obtained at two different temperatures, $\beta^{(1)}\Delta=100$ and $\beta^{(2)}\Delta=10$ respectively, in TS\protect\nobreakdash-TS (top) and S-S (bottom) junctions, and for several transparencies.}\label{fig_noise_temp}
\end{figure}
\begin{figure}
\includegraphics[scale=0.6]{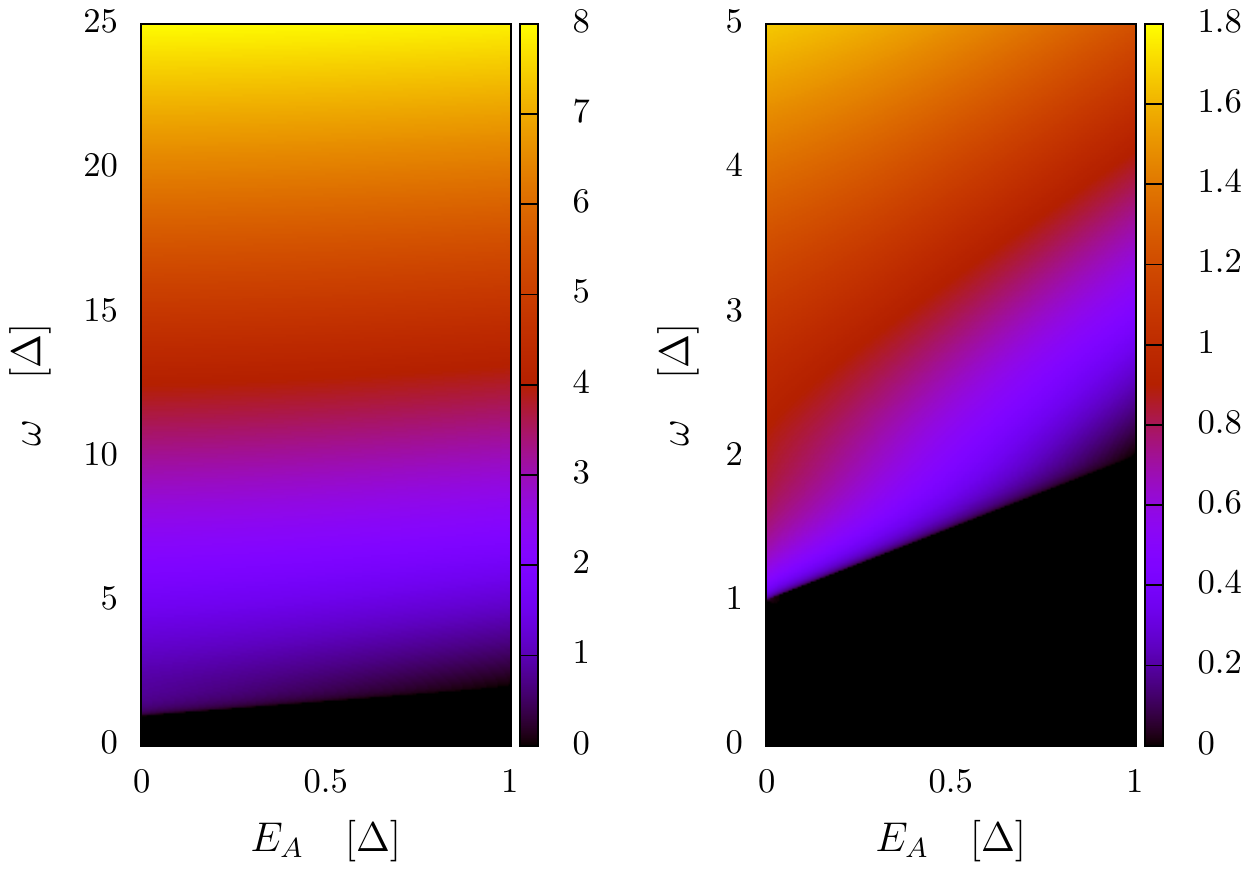}
\caption{Non-resonant noise $S_\text{nr}/T$ in a TS-TS junction, as a function of $E_A$ and $\omega$. The first panel gives the map on a large scale of energies while the second panel is an enlargement around the gap region.}\label{fig_noise2}
\end{figure}

Let us briefly investigate finite temperature effects by increasing temperature $\beta^{-1}=0.1\Delta$. The main change is the appearance of noise for energies ranging between the curves $\Delta-E_A(\varphi)$ and $\Delta+E_A(\varphi)$, as illustrated in Fig.~\ref{fig_noise_temp}. More interestingly, let us also mention the noticeable differences in a TS-TS junction just above the gap and around $\varphi=\pi$ (zero of the Andreev energy) where finite values of $\beta$ cannot reproduce the zero-temperature limit.

A final remark is worthy. In the expression of the non-resonant noise $S_\text{nr}$ in a TS-TS junction, the transparency appears as an overall factor and in the Andreev energy expression so that dependences on $\varphi$ and $T$ for $S_\text{nr}/T$ can be recast in the Andreev energy dependence (contrary to the case of a S-S junction where $\rho(\varphi)$ given in Eq.~\eqref{eq_rhofunc} provides another dependence on $T$). In Fig.~\ref{fig_noise2}, the map in the $(E_A,\omega)$ plane is displayed. Remark that, for a given transparency $T$, Andreev energies range in $[0,\sqrt{T}]$.

\section{Conclusion}\label{sec_conclusion}

More and more convincing signatures compatible with the presence of Majorana fermions in 1D semiconducting wires have been accumulated with recent observations of quantized zero-bias conductance~\cite{prl119_nichele,nature556_zhang}. Despite the fact that the confining potential has to be carefully settled in order to avoid spurious non-topological subgap states~\cite{prb97_stanescu}, 2D layouts of 1D $p$\nobreakdash-wave superconductors~\cite{nphys_alicea} are still serious candidates to provide the exchange statistics decisive evidence for Majorana fermions. Before the implementation of these networks, quantum transport in 1D systems supporting Majorana fermions, e.g. the junction between 1D topological superconducting wires investigated in this article, deserves deeper study. 

We have presented a unified description of S-S and TS-TS junctions in terms of scattering eigenstates. We have shown how Majorana fermions emerge in a TS-TS junction with phase difference $\varphi=\pi$ by explicitely writting the proper linear combinations of (degenerate) zero-energy Andreev bound states. The Josephson current carried by Andreev states can be written in a unified way as the derivative of the Andreev energy $\left\langle I\right\rangle\propto\frac{\text{d}E_A}{\text{d}\varphi}$. Transitions inside the Andreev sector are noiseless in the topological case since Andreev states are current eigenstates (a result which is closely related to the emergence of Majorana fermions) in contrast with the BCS case for which a noise resonance at energy $\omega=2E_A$ is expected for $T<1$ (finite backscattering). While resonant zero frequency noise is suppressed by lowering temperature in a S-S junction, it is expected to persist in a TS-TS junction operating with $\varphi=\pi$. Transitions which involve continuum states give rise to a non-resonant noise which has been computed. For low temperatures, it exhibits a $\Delta+E_A$ frequency threshold due to AC transitions. CC transitions which occur for energies larger than $2\Delta$ impose an asymptotic linear behavior. A detailed comparison between the two types of junction has been carried out and characteristic features in a TS-TS junction have been highlighted, some of them related to the existence of zero-energy modes in this topologically nontrivial junction. 

In the linear response framework, one can adress the issue of computing the current susceptibility which is closely related to the noise $S(\omega)$ studied in this article. Indeed, the imaginary part of the current susceptibility, which gives the linear absorption rate, is proportional to the difference $S(\omega)-S(-\omega)$. In a S-S junction our results coincide with those of Ref.~\onlinecite{prb87_kos}. More interestingly, the noise calculations can be straightforwardly used for computing the current susceptibility in a TS-TS junction. The characteritic features detailed in the main text of this article for the noise will be recovered for the imaginary part of the susceptibility. Let us mention that dynamic current susceptibility has recently been proposed as a probing tool for the presence of Majorana bound states in a superconduting ring geometry~\cite{prb97_bouchiat}. One can also adress the study of Andreev level qubit~\cite{prl90_zazunov} population dynamics. In a conventional Andreev qubit, long-lived quasiparticles can be trapped~\cite{prl106_zgirski}, a phenomenon known as quasiparticle poisoning. Within our framework, using the calculations of current matrix elements, one can compare the transition rates between quasiparticle states due to the coupling of the junction to its environment (external circuit, phonons) calculated in Ref.~\onlinecite{prb89_olivares}, with those calculated in a topological Andreev qubit. When considered as a function of Andreev energy, the latter are simply given by the $T\to1$ limiting case given in Ref.~\onlinecite{prb89_olivares} renormalized by an overall multiplication with $T/2$. Indeed since $d_\star S_\text{nr}$ in both S-S and TS-TS junctions coincide in the limit $T=1$ and since $\left.(S_\text{nr}/T)\right|_\text{TS-TS}$ considered as function of $E_A$ and $\omega$ does not depend anymore on $T$, we get
\begin{align}
\left.S_\text{nr}(E_A,\omega)\right|_\text{TS-TS}^T&=T\left.S_\text{nr}(E_A,\omega)\right|_\text{TS-TS}^{T=1}\notag\\
&=\frac{T}{2}\,\left.S_\text{nr}(E_A,\omega)\right|_\text{S-S}^{T=1}~.
\end{align}
The same argument holds for the squared current matrix elements involved in the transition rates of Ref~\onlinecite{prb89_olivares}. Stationary occupation probabilities can be computed by adapting the calculation of Ref.~\onlinecite{prb90_zazunov}.   



\acknowledgements
We acknowledge the support of the French National
Research Agency, through the project ANR NanoQuartets
(ANR-12-BS1000701). Part of this work
has been carried out in the framework of the Labex
Archim\`ede ANR-11-LABX-0033. The project leading to this publication has received funding from Excellence Initiative of Aix-Marseille University - A*MIDEX, a French ``Investissements d'Avenir'' programme.

\appendix

\section{BdG scattering eigenstates}\label{app_eigen}

A continuum state can be written as a sum of incoming and outgoing waves according to $\chi_p=\chi_p^\text{in}+\chi_p^\text{out}$. If we define $\eta_E=\text{sign}(E)$, and the angle $\theta_E$ such that $\cosh\theta_E=|E|/\Delta$, $\sinh\theta_E=\sqrt{\left(E/\Delta\right)^2-1}=\eta_E k_E\xi_0$, the associated bispinors read
\begin{align}
\chi_{p=(E,s)}^\text{in}(x)=&\Theta(-x)\,\frac{\text{e}^{ik_Ex}}{\sqrt{l}}
\left(\begin{matrix}
\delta_{s,1}[\chi_e]_E\\
\delta_{s,2}[\chi_h]_E
\end{matrix}\right)
\notag\\
+&\Theta(x)\,\frac{\text{e}^{-ik_Ex}}{\sqrt{l}}
\left(\begin{matrix}
\delta_{s,4}[\chi_h]_E\\
\delta_{s,3}[\chi_e]_E
\end{matrix}\right)~,
\\
\chi_{p=(E,s)}^\text{out}(x)=&\Theta(-x)\,\frac{\text{e}^{-ik_Ex}}{\sqrt{l}}
\left(\begin{matrix}
a_p[\chi_h]_E\\
b_p[\chi_e]_E
\end{matrix}\right)
\notag\\
+&\Theta(x)\,\frac{\text{e}^{ik_Ex}}{\sqrt{l}}
\left(\begin{matrix}
c_p[\chi_e]_E\\
d_p[\chi_h]_E
\end{matrix}\right)~,
\label{continuum_wf}
\end{align}
with the following spinors which describe electron-like or hole-like excitations 
\be
[\chi_{e,h}]_E=\frac{\text{e}^{\pm\frac{\theta_E}{2}\tau_z}}{\sqrt{2\cosh\theta_E}}
\left(
\begin{matrix}
1\\
\eta_E
\end{matrix}
\right)~,
\ee
and where $l>\!\!>\xi_0$ is the wire length. Proceeding in the same way for subgap states, we define $\sigma_E=\text{sign}(E)$, $\cos\gamma_E=|E|/\Delta$, $\sin\gamma_E=\sigma_E\sqrt{1-\left(E/\Delta\right)^2}=\sigma_E \kappa_E\xi_0$. Subgap bound states can be written as
\begin{equation}
\chi_E(x)=\frac{\text{e}^{-\kappa_E|x|}}{\sqrt{\xi_0}}
\left[
\Theta(-x)
\left(
\begin{matrix}
a_E[\tilde{\chi}_h]_E\\
b_E[\tilde{\chi}_e]_E
\end{matrix}
\right)
+\Theta(x)
\left(
\begin{matrix}
c_E[\tilde{\chi}_e]_E\\
d_E[\tilde{\chi}_h]_E
\end{matrix}
\right)
\right]
\ee
with
\be
[\tilde{\chi}_{e,h}]_E=\frac{\text{e}^{\pm\frac{i\gamma_E}{2}\tau_z}}{\sqrt{2}}
\left(
\begin{matrix}
1\\
\sigma_E
\end{matrix}
\right)~.\label{subgap_wf}
\end{equation}
The matching condition~\eqref{eq_matching} written for continuum wavefunctions~\eqref{continuum_wf} leads to symmetries between the coefficients of different scattering states $s$ which are given in the following table
\renewcommand{\arraystretch}{1.5}
\begin{table}[H]
\centering
\begin{tabular}{c|c|c|c|c|}
 & $s=1$ & $s=2$ & $s=3$ & $s=4$\\ \hline
$a_{(E,s)}(\varphi)$ & $A(\theta_E,\varphi)$ & $B(-\theta_E,\varphi)$ & $-D(\theta_E,-\varphi)$ & $C(-\theta_E,-\varphi)$\\ \hline
$b_{(E,s)}(\varphi)$ & $B(\theta_E,\varphi)$ & $A(-\theta_E,\varphi)$ & $C(\theta_E,-\varphi)$ & $-D(-\theta_E,-\varphi)$\\ \hline
$c_{(E,s)}(\varphi)$ & $C(\theta_E,\varphi)$ & $D(-\theta_E,\varphi)$ & $-B(\theta_E,-\varphi)$ & $A(-\theta_E,-\varphi)$\\ \hline
$d_{(E,s)}(\varphi)$ & $D(\theta_E,\varphi)$ & $C(-\theta_E,\varphi)$ & $A(\theta_E,-\varphi)$ & $-B(-\theta_E,-\varphi)$\\
\hline
\end{tabular}\caption{}
\end{table}
\renewcommand{\arraystretch}{1}
\noindent One can easily prove that this table constitutes a unitary matrix. In particular, because of the orthonormality between the columns, the scattering states $(E,s)$ for $s=1..4$ form a basis of a given energy $E$ subspace. The four functions $A,B,C,D$ are given by the resolution of the linear system~\eqref{eq_matching} for $s=1$. If we define $Q(\theta,\varphi)=g^2(\theta)-Tf^2(\varphi)$ and $\tilde{X}=QX$ for $X=A,B,C,D$, we have
\begin{subequations}
\begin{align}
&\tilde{A}(\theta,\varphi)=-\left[\cosh\theta\,\kappa^2(\varphi)\xi_0^2+i\sinh\theta\,E_A(\varphi)\delta_A(\varphi)\Delta^{-2}\right]~,\\
&\tilde{B}(\theta,\varphi)=\sqrt{(1-T)}\sinh\theta\,\,g(\theta)~,\\
&\tilde{C}(\theta,\varphi)=\sqrt{T}\sinh\theta\sinh\left(\theta-i\frac{\varphi}{2}\right)~,\\
&\tilde{D}(\theta,\varphi)=\sqrt{T(1-T)}\sinh\theta\,f(\varphi)~.
\end{align}
\end{subequations}
The matching condition~\eqref{eq_matching} written for subgap wavefunctions~\eqref{subgap_wf} provides a homogeneous linear system for $a_E,b_E,c_E,d_E$ coefficients which leads to the quantization of the energy: the so-called Andreev levels have opposite energies $E_\sigma=\sigma E_A$ with $\sigma=\pm$ and where the Andreev energy $E_A$ is given in Eq.~\eqref{eq_andreev_energy}. The coefficients $a_\sigma, b_\sigma, c_\sigma, d_\sigma$ (quantities which were labeled with $E_\sigma$ appear now with an index $\sigma$) can be conveniently expressed thanks to the quantity ${\cal R}_\sigma=\xi_0\kappa-\sigma\,\Delta^{-1}\delta_A$, where $\kappa$ and $\delta_A$ are defined in Eqs.~\eqref{eq_andreev_kappa} and~\eqref{eq_andreev_delta} respectively. They are given in the table below
\renewcommand{\arraystretch}{1.5}
\begin{table}[H]
\centering
\begin{tabular}{c|c|c|}
 & topological & BCS \\ \hline
$s(\varphi)$ & $\text{sign}\left[\cos\frac{\varphi}{2}\right]$ & $\text{sign}\left[\sin\frac{\varphi}{2}\right]$\\ \hline
$a_\sigma(\varphi)$ & $s(\varphi)\sqrt{\frac{{\cal R}_\sigma(\varphi)}{2}}$ & $-\sigma\,s(\varphi)\sqrt{\frac{{\cal R}_\sigma(\varphi)}{2}}$\\ \hline
$b_\sigma(\varphi)$ & $-i\sigma\,s(\varphi)\sqrt{\frac{{\cal R}_{-\sigma}(\varphi)}{2}}$ & $-\sigma\,s(\varphi)\sqrt{\frac{{\cal R}_{-\sigma}(\varphi)}{2}}$\\ \hline
$c_\sigma(\varphi)$ & $\sqrt{\frac{{\cal R}_\sigma(\varphi)}{2}}$ & $\sqrt{\frac{{\cal R}_\sigma(\varphi)}{2}}$\\ \hline
$d_\sigma(\varphi)$ & $-i\sigma\sqrt{\frac{{\cal R}_{-\sigma}(\varphi)}{2}}$ & $-\sqrt{\frac{{\cal R}_{-\sigma}(\varphi)}{2}}$\\
\hline
\end{tabular}\caption{}
\end{table}
\renewcommand{\arraystretch}{1}
\noindent Notice the important relation which holds for topological Andreev states
\be
a_\sigma^\ast a_{-\sigma}+b_\sigma^\ast b_{-\sigma}=c_\sigma^\ast c_{-\sigma}+d_\sigma^\ast d_{-\sigma}=0~,
\ee
at the origin of the construction of Majorana wavefunctions (because it yields $\left[C\chi_\sigma\right]^\ast\propto\chi_{-\sigma}$) and of the cancellation of the noise due to Andreev transitions: $S_{AA}=0$.

\bibliographystyle{unsrtnat}
\bibliography{biblio}

\begin{thebibliography}{43}
\providecommand{\natexlab}[1]{#1}
\providecommand{\url}[1]{\texttt{#1}}
\expandafter\ifx\csname urlstyle\endcsname\relax
  \providecommand{\doi}[1]{doi: #1}\else
  \providecommand{\doi}{doi: \begingroup \urlstyle{rm}\Url}\fi

\bibitem[Majorana(1937)]{majorana}
E.~Majorana.
\newblock \emph{Nuovo Cim}, 14:\penalty0 171, 1937.

\bibitem[Elliott and Franz(2015)]{rmp87_elliott}
S.~R. Elliott and M.~Franz.
\newblock \emph{Rev. Mod. Phys.}, 87:\penalty0 137, 2015.

\bibitem[Read and Green(2000)]{prb61_read}
N.~Read and D.~Green.
\newblock \emph{Phys. Rev. B}, 61:\penalty0 10267, 2000.

\bibitem[Kitaev(2001)]{kitaev01}
A.~Y. Kitaev.
\newblock \emph{Physics-Uspekhi}, 44:\penalty0 131, 2001.

\bibitem[Fu and Kane(2008)]{prl100_fu_kane}
L.~Fu and C.~L. Kane.
\newblock \emph{Phys. Rev. Lett.}, 100:\penalty0 096407, 2008.

\bibitem[Alicea(2012)]{alicea_rev}
J.~Alicea.
\newblock \emph{Reports on Progress in Physics}, 75:\penalty0 076501, 2012.

\bibitem[Beenakker(2013)]{beenakker_rev}
C.W.J. Beenakker.
\newblock \emph{Annual Review of Condensed Matter Physics}, 4:\penalty0 113,
  2013.

\bibitem[Leijnse and Flensberg(2012)]{flensberg_rev}
M.~Leijnse and K.~Flensberg.
\newblock \emph{Semiconductor Science and Technology}, 27:\penalty0 124003,
  2012.

\bibitem[Ivanov(2001)]{prl86_ivanov}
D.~A. Ivanov.
\newblock \emph{Phys. Rev. Lett.}, 86:\penalty0 268, 2001.

\bibitem[Nayak et~al.(2008)Nayak, Simon, Stern, Freedman, and
  Das~Sarma]{rmp80_nayak}
C.~Nayak, S.~H. Simon, A.~Stern, M.~Freedman, and S.~Das~Sarma.
\newblock \emph{Rev. Mod. Phys.}, 80:\penalty0 1083, 2008.

\bibitem[Kitaev(2003)]{ann_phys_kitaev}
A.~Y. Kitaev.
\newblock \emph{Annals of Physics}, 303:\penalty0 2, 2003.

\bibitem[Sau et~al.(2010)Sau, Lutchyn, Tewari, and Das~Sarma]{prl104_sau}
J.~D. Sau, R.~M. Lutchyn, S.~Tewari, and S.~Das~Sarma.
\newblock \emph{Phys. Rev. Lett.}, 104:\penalty0 040502, 2010.

\bibitem[Alicea(2010)]{prb81_alicea}
J.~Alicea.
\newblock \emph{Phys. Rev. B}, 81:\penalty0 125318, 2010.

\bibitem[Lutchyn et~al.(2010)Lutchyn, Sau, and Das~Sarma]{prl105_lutchyn}
R.~M. Lutchyn, J.~D. Sau, and S.~Das~Sarma.
\newblock \emph{Phys. Rev. Lett.}, 105:\penalty0 077001, 2010.

\bibitem[Oreg et~al.(2010)Oreg, Refael, and von Oppen]{prl105_oreg}
Y.~Oreg, G.~Refael, and F.~von Oppen.
\newblock \emph{Phys. Rev. Lett.}, 105:\penalty0 177002, 2010.

\bibitem[Alicea et~al.(2011)Alicea, Oreg, Refael, von Oppen, and
  Fisher]{nphys_alicea}
J.~Alicea, Y.~Oreg, G.~Refael, F.~von Oppen, and M.~P.~A. Fisher.
\newblock \emph{Nature Physics}, 7:\penalty0 412, 2011.

\bibitem[Mourik et~al.(2012)Mourik, Zuo, Frolov, Plissard, Bakkers, and
  Kouwenhoven]{science336_mourik}
V.~Mourik, K.~Zuo, S.~M. Frolov, S.~R. Plissard, E.~P. A.~M. Bakkers, and L.~P.
  Kouwenhoven.
\newblock \emph{Science}, 336:\penalty0 1003, 2012.

\bibitem[Das et~al.(2012)Das, Ronen, Most, Oreg, Heiblum, and
  Shtrikman]{nat_phys_das}
A.~Das, Y.~Ronen, Y.~Most, Y.~Oreg, M.~Heiblum, and H.~Shtrikman.
\newblock \emph{Nature Physics}, 8:\penalty0 887, 2012.

\bibitem[Churchill et~al.(2013)Churchill, Fatemi, Grove-Rasmussen, Deng,
  Caroff, Xu, and Marcus]{prb87_churchill}
H.~O.~H. Churchill, V.~Fatemi, K.~Grove-Rasmussen, M.~T. Deng, P.~Caroff, H.~Q.
  Xu, and C.~M. Marcus.
\newblock \emph{Phys. Rev. B}, 87:\penalty0 241401, 2013.

\bibitem[Deng et~al.(2012)Deng, Yu, Huang, Larsson, Caroff, and
  Xu]{nano_letters_deng}
M.~T. Deng, C.~L. Yu, G.~Y. Huang, M.~Larsson, P.~Caroff, and H.~Q. Xu.
\newblock \emph{Nano Letters}, 12:\penalty0 6414, 2012.

\bibitem[Finck et~al.(2013)Finck, Van~Harlingen, Mohseni, Jung, and
  Li]{prl110_finck}
A.~D.~K. Finck, D.~J. Van~Harlingen, P.~K. Mohseni, K.~Jung, and X.~Li.
\newblock \emph{Phys. Rev. Lett.}, 110:\penalty0 126406, 2013.

\bibitem[Fu and Kane(2009)]{prb79_fu_kane}
L.~Fu and C.~L. Kane.
\newblock \emph{Phys. Rev. B}, 79:\penalty0 161408, 2009.

\bibitem[Rokhinson et~al.(2012)Rokhinson, Liu, and Furdyna]{nature_rokhinson}
L.~P. Rokhinson, X.~Liu, and J.~K. Furdyna.
\newblock \emph{Nature Physics}, 8:\penalty0 795, 2012.

\bibitem[Chiu et~al.(2018)Chiu, Sau, and Das~Sarma]{prb97_sau}
C.-K. Chiu, J.~D. Sau, and S.~Das~Sarma.
\newblock \emph{Phys. Rev. B}, 97:\penalty0 035310, 2018.

\bibitem[Hansen et~al.(2018)Hansen, Danon, and Flensberg]{prb97_flensberg}
E.~B. Hansen, J.~Danon, and K.~Flensberg.
\newblock \emph{Phys. Rev. B}, 97:\penalty0 041411, 2018.

\bibitem[Albrecht et~al.(2016)Albrecht, Higginbotham, Madsen, Kuemmeth,
  Jespersen, Nygård, Krogstrup, and Marcus]{nature_albrecht}
S.~M. Albrecht, A.~P. Higginbotham, M.~Madsen, F.~Kuemmeth, T.~S. Jespersen,
  J.~Nygård, P.~Krogstrup, and C.~M. Marcus.
\newblock \emph{Nature}, 531:\penalty0 206, 2016.

\bibitem[Nichele et~al.(2017)Nichele, Drachmann, Whiticar, O'Farrell, Suominen,
  Fornieri, Wang, Gardner, Thomas, Hatke, Krogstrup, Manfra, Flensberg, and
  Marcus]{prl119_nichele}
F.~Nichele, A.~C.~C. Drachmann, A.~M. Whiticar, E.~C.~T. O'Farrell, H.~J.
  Suominen, A.~Fornieri, T.~Wang, G.~C. Gardner, C.~Thomas, A.~T. Hatke,
  P.~Krogstrup, M.~J. Manfra, K.~Flensberg, and C.~M. Marcus.
\newblock \emph{Phys. Rev. Lett.}, 119:\penalty0 136803, 2017.

\bibitem[Zhang et~al.(2018)Zhang, Liu, Gazibegovic, Xu, Logan, Wang, van Loo,
  Bommer, de~Moor, Car, Op~het Veld, van Veldhoven, Koelling, Verheijen,
  Pendharkar, Pennachio, Shojaei, Lee, Palmstrøm, Bakkers, Sarma, and
  Kouwenhoven]{nature556_zhang}
H.~Zhang, C.-X. Liu, S.~Gazibegovic, D.~Xu, J.~A. Logan, G.~Wang, N.~van Loo,
  J.~D.~S. Bommer, M.~W.~A. de~Moor, D.~Car, R.~L.~M. Op~het Veld, P.~J. van
  Veldhoven, S.~Koelling, M.~A. Verheijen, M.~Pendharkar, D.~J. Pennachio,
  B.~Shojaei, J.~S. Lee, C.~J. Palmstrøm, E.~P. A.~M. Bakkers, S.~Das Sarma,
  and L.~P. Kouwenhoven.
\newblock \emph{Nature}, 556:\penalty0 74, 2018.

\bibitem[Furusaki and Tsukada(1990)]{phyb165}
A.~Furusaki and M.~Tsukada.
\newblock \emph{Physica B}, 165-166:\penalty0 967, 1990.

\bibitem[Beenakker and van Houten(1991)]{prl66_beenakker}
C.~W.~J. Beenakker and H.~van Houten.
\newblock \emph{Phys. Rev. Lett.}, 66:\penalty0 3056, 1991.

\bibitem[Beenakker(1991)]{prl67_beenakker}
C.~W.~J. Beenakker.
\newblock \emph{Phys. Rev. Lett.}, 67:\penalty0 3836, 1991.

\bibitem[Mart\'{\i}n-Rodero et~al.(1996)Mart\'{\i}n-Rodero, Yeyati, and
  Garc\'{\i}a-Vidal]{prb53_martin-rodero}
A.~Mart\'{\i}n-Rodero, A.~Levy Yeyati, and F.~J. Garc\'{\i}a-Vidal.
\newblock \emph{Phys. Rev. B}, 53:\penalty0 R8891, 1996.

\bibitem[Shumeiko et~al.(1993)Shumeiko, Wendin, and Bratus']{prb48_shumeiko}
V.~S. Shumeiko, G.~Wendin, and E.~N. Bratus'.
\newblock \emph{Phys. Rev. B}, 48:\penalty0 13129, 1993.

\bibitem[Zazunov et~al.(2003)Zazunov, Shumeiko, Bratus', Lantz, and
  Wendin]{prl90_zazunov}
A.~Zazunov, V.~S. Shumeiko, E.~N. Bratus', J.~Lantz, and G.~Wendin.
\newblock \emph{Phys. Rev. Lett.}, 90:\penalty0 087003, 2003.

\bibitem[Zazunov et~al.(2005)Zazunov, Shumeiko, Wendin, and
  Bratus']{prb71_zazunov}
A.~Zazunov, V.~S. Shumeiko, G.~Wendin, and E.~N. Bratus'.
\newblock \emph{Phys. Rev. B}, 71:\penalty0 214505, 2005.

\bibitem[Olivares et~al.(2014)Olivares, Yeyati, Bretheau, Girit, Pothier, and
  Urbina]{prb89_olivares}
D.~G. Olivares, A.~Levy Yeyati, L.~Bretheau, \ifmmode \mbox{\c{C}}\else
  \c{C}\fi{}.~\"O. Girit, H.~Pothier, and C.~Urbina.
\newblock \emph{Phys. Rev. B}, 89:\penalty0 104504, 2014.

\bibitem[Zazunov et~al.(2014)Zazunov, Brunetti, Yeyati, and
  Egger]{prb90_zazunov}
A.~Zazunov, A.~Brunetti, A.~Levy Yeyati, and R.~Egger.
\newblock \emph{Phys. Rev. B}, 90:\penalty0 104508, 2014.

\bibitem[Riwar et~al.(2015)Riwar, Houzet, Meyer, and Nazarov]{condmat_riwar}
R.-P. Riwar, M.~Houzet, J.~S. Meyer, and Y.~V. Nazarov.
\newblock \emph{Journal of Physics: Condensed Matter}, 27:\penalty0 095701,
  2015.

\bibitem[Zazunov et~al.(2016)Zazunov, Egger, and Levy~Yeyati]{prb94_zazunov}
A.~Zazunov, R.~Egger, and A.~Levy~Yeyati.
\newblock \emph{Phys. Rev. B}, 94:\penalty0 014502, 2016.

\bibitem[Stanescu and Das~Sarma(2018)]{prb97_stanescu}
T.~D. Stanescu and S.~Das~Sarma.
\newblock \emph{Phys. Rev. B}, 97:\penalty0 045410, 2018.

\bibitem[Kos et~al.(2013)Kos, Nigg, and Glazman]{prb87_kos}
F.~Kos, S.~E. Nigg, and L.~I. Glazman.
\newblock \emph{Phys. Rev. B}, 87:\penalty0 174521, 2013.

\bibitem[Trif et~al.(2018)Trif, Dmytruk, Bouchiat, Aguado, and
  Simon]{prb97_bouchiat}
M.~Trif, O.~Dmytruk, H.~Bouchiat, R.~Aguado, and P.~Simon.
\newblock \emph{Phys. Rev. B}, 97:\penalty0 041415, 2018.

\bibitem[Zgirski et~al.(2011)Zgirski, Bretheau, Le~Masne, Pothier, Esteve, and
  Urbina]{prl106_zgirski}
M.~Zgirski, L.~Bretheau, Q.~Le~Masne, H.~Pothier, D.~Esteve, and C.~Urbina.
\newblock \emph{Phys. Rev. Lett.}, 106:\penalty0 257003, 2011.

\end{thebibliography}

\end{document}